\documentclass[11pt, a4paper,twoside]{article}
\usepackage{trinhcokenofont}

\usepackage{graphicx}
\usepackage{amsmath, amssymb, amsthm, cleveref}
\usepackage[inactive,tightpage]{preview}

\usepackage[normalem]{ulem}
\usepackage{mathtools}
\usepackage{esvect}

\usepackage[outline]{contour}

\usepackage{enumitem}

\usepackage{tocloft}
\pdfmapfile{+rsfso.map}
\DeclareSymbolFont{rsfso}{U}{rsfso}{m}{n}
\DeclareSymbolFontAlphabet{\mathscr}{rsfso}

\usepackage{placeins}
\usepackage{afterpage}

\usepackage{bm}
\usepackage{microtype}

\usepackage{hyperref}
\usepackage[usenames, dvipsnames]{xcolor}
\hypersetup{colorlinks=true,linkcolor=MidnightBlue,citecolor=MidnightBlue}

\usepackage{natbib}
\usepackage{array}
\usepackage{booktabs}
\usepackage{multirow}

\usepackage{subcaption}
\usepackage{tabularx}
\newcolumntype{Y}{>{\centering\arraybackslash}X}
\usepackage{wrapfig,lipsum}

\usepackage[percent]{overpic}
\usepackage[ruled,linesnumbered]{algorithm2e}
\usepackage{float}

\usepackage{tikz}
\usetikzlibrary{arrows.meta}

\newcommand*{\ep}{\epsilon}

\renewcommand*{\i}{\mathrm{i}}
\newcommand*{\im}{\mathrm{i}}
\newcommand*{\e}{\mathrm{e}}
\newcommand*{\Oh}{\mathcal{O}}
\renewcommand*{\Re}{\operatorname{Re}}
\renewcommand*{\Im}{\operatorname{Im}}
\newcommand*{\Arg}{\operatorname{Arg}}

\newcommand{\ql}{\overline{q}}
\newcommand*{\Ei}{\text{Ei}}

\newcommand*{\sech}{\mathrm{sech}}

\newcommand*{\de}{\operatorname{d\!}{}} 
\newcommand{\dd}[2]{\frac{\de#1}{\de#2}}

\usepackage{mathabx}
\usepackage{esint} 
\def\Xint#1{\mathchoice
   {\XXint\displaystyle\textstyle{#1}}%
   {\XXint\textstyle\scriptstyle{#1}}%
   {\XXint\scriptstyle\scriptscriptstyle{#1}}%
   {\XXint\scriptscriptstyle\scriptscriptstyle{#1}}%
   \!\int}
\def\XXint#1#2#3{{\setbox0=\hbox{$#1{#2#3}{\int}$}
     \vcenter{\hbox{$#2#3$}}\kern-.5\wd0}}

\def\YYint#1#2#3{{\setbox0=\hbox{$#1{#2#3}{\int}$}
     \vcenter{\hbox{\scalebox{1}[-1]{$#2#3$}}}\kern-.5\wd0}}

\def\dashint{\Xint-}

\newcommand{\intleft}{\Xint\curvearrowleft}

\usepackage{mdframed}

\title{Pathologies of low-Froude free-surface flows over smoothed bodies}

\author{PHT and YJ-L}

\theoremstyle{definition}
\newtheorem{remark}{Remark}

\begin{document}


\begin{center}
	\Large \scshape\MakeLowercase{Pathologies in the asymptotics of low-Froude \\ free-surface waves over smooth bodies}
\end{center}

\vspace*{0.5\baselineskip}

\begin{center}
\noindent {\fontsize{12}{16}\selectfont\scshape{By Yyanis Johnson-Llambias \& Philippe H. Trinh}} \\

\emph{Department of Mathematical Sciences, University of Bath, BA2 7AY} \\
 
\vspace*{0.5\baselineskip}
\emph{A paper dedicated to the $70^{\mathrm{th}}$ birthday of Prof. J.-M. Vanden-Broeck} \\
\emph{(Submitted manuscript)}
\end{center}


\begin{abstract}
\noindent In the study of low-speed or low-Froude flows of a potential gravity-driven fluid past a wave-generating object, the traditional asymptotic expansion in powers of the Froude number predicts a waveless free-surface at every order. This is due to the fact that the waves are, in fact, exponentially small and beyond-all-orders of the naive expansion. The theory of exponential asymptotics indicates that such exponentially-small water waves are switched-on across so-called Stokes lines---these curves partition the fluid-domain into wave-free regions and regions with waves. In prior studies, Stokes lines are associated with singularities in the flow field such as stagnation points, or corners in submerged objects or rough beds. In this work, we present a smoothed geometry that was recently highlighted by Pethiyagoda \emph{et al.} [{\itshape Int. J. Numer. Meth. Fluids}. 2018; \textbf{86}:607--624] as capable of producing waves, yet paradoxically exhibiting no obvious Stokes line insofar as conventional exponential asymptotics theory. In this work, we demonstrate that the Stokes line for this smooth geometry originates from an essential singularity at infinity in the analytic continuation of free-surface quantities. We discuss some of the difficulties in extending the typical methodology of exponential asymptotics to general wave-structure interaction problems with smooth geometries.
\end{abstract}

\section{Introduction}

\noindent The motivation of this work stems from an apparent violation of a criterion originally proposed by \cite{chapmanExponentialAsymptoticsGravity2006} for the existence of small-scale gravity-driven free-surface waves at low speeds due to a disturbance in the flow. As was originally noted by \cite{ogilvieWaveResistanceLow1968}, in the low-speed (or low-Froude) limit, free-surface waves are exponentially small in the Froude number and hence {\itshape beyond-all-orders} of a naive asymptotic expansion in algebraic powers. Their development requires techniques in exponential asymptotics. Such a framework was developed through numerous authors such as \cite{vb_1977,ogilvieWaterWavesGenerated1982,brandsma_1985a,tulinExactTheoryGravity1982}. For this specific case of wave-structure-driven water waves at low speeds, the modern incarnation of the exponential asymptotics methodology was advanced by \cite{chapmanExponentialAsymptoticsCapillary2002} and \cite{chapmanExponentialAsymptoticsGravity2006} for the respective case of surface capillary waves and surface gravity waves. In order to introduce the pathology to be discussed in this work, we first require a review of the specific wave-making criterion that was developed in this last work by \cite{chapmanExponentialAsymptoticsGravity2006}.




Consider two-dimensional potential (inviscid and irrotational) gravity-driven free-surface flow past an object. The object can be surface-piercing or submerged. Simple classical geometries include the case of a semi-infinite ship's stern~\citep{vb_1977}, a rectangular step in a channel~\citep{kingFreesurfaceFlowStream1990}, and a semi-circular cylinder~\citep{forbesFreesurfaceFlowSemicircular1982}. Typically, the task is to solve for the free-surface unknowns, say the streamline speed $q$ and angle $\theta$, as a function of the velocity potential, $\phi$. Mathematically, the simplest possible formulations take the form of Bernoulli's equation and a boundary integral that imposes a relationship between $q$ and $\theta$. Other water-wave formulations are possible. 

As a particular example, consider the case of flow in a channel with a bottom topography. The upstream flow speed is $U$, the upstream channel height is $L/\pi$, and gravity is $g$. The problem is non-dimensionalised so that the flow is contained within a strip in the complex potential plane, $w = \phi + \im \psi$, between $\psi = 0$ (free surface) and $\psi = -\pi$ (channel bottom). The strip in the $w$-plane can then be mapped to the upper half-$\zeta$-plane via $\zeta = \xi + \im \eta = \e^{-w}$. Then the two governing equations, applied to the free surface $\xi \geq 0$ or $-\infty < \phi < \infty$ are \citep[\S{2}]{chapmanExponentialAsymptoticsGravity2006}:
\begin{subequations} \label{eq:govset}
	\begin{gather}
	\ep q^2 \dd{q}{\phi} =  -\sin\theta, \label{eq:mybern} \\
	\log q(\xi) = \underbrace{-\frac{1}{\pi}\int_{-\infty}^{0}\frac{\theta(\xi^\prime)}{\xi^\prime-\xi}\de \xi^\prime}_{\log q_s} -\frac{1}{\pi}\dashint_{0}^{\infty}\frac{\theta(\xi^\prime)}{\xi^\prime-\xi}\de \xi^\prime. \label{govset_int}
	\end{gather}
\end{subequations}
where $\epsilon = U^2/(gL)$ is the square of the Froude number, characterising the balance between inertial and gravitational effects. We have defined the function $q_s = q_s(\xi)$ above so that it may be referenced later---it can be interpreted as a `shape' function.

The bottom topography can then be specified via the angle $\theta$ for $\xi$ along the negative real axis. For example, in the case of flow past an inlined step with angle of inclination $\pi\sigma$, $0 < \sigma < 1$, we choose $a$ and $b$ such that $0 < a < b$. Then set $\theta = 0$ for $\xi \in (-\infty, -b) \cup (-a, 0)$ and $\theta = \pi\sigma$ for $\xi \in (-b, -a)$. Computing the first integral in \eqref{govset_int} yields
\begin{equation} \label{eq:q0step}
q_s = \left( \frac{\xi + b}{\xi + a}\right)^{\sigma} = \left( \frac{\e^{-\phi} + b}{\e^{-\phi} + a}\right)^{\sigma}.
\end{equation}
For instance, with $\sigma = 1/2$, this is the rectangular step in a channel. Similarly, the function 
\begin{equation} \label{eq:q0tri}
q_s = \left[\frac{(\xi+1)(\xi+b)}{(\xi+c)^2}\right]^{\sigma}
= \left[\frac{(\e^{-\phi}+1)(\e^{-\phi}+b)}{(\e^{-\phi}+c)^2}\right]^{\sigma},
\end{equation}
corresponds to flow over a triangular obstruction with initial inclination $\sigma \pi$, and with $1 < c < b$. The derivation of these and similar quantities is reviewed in \cref{sec:2dbump_formulation}.

\begin{remark} \label{rem:q0}
Note that  $q_s$ corresponds to the leading-order speed in the limit of small Froude numbers. Thus $q \sim q_0 = q_s$ and $\theta \sim \theta_0 \equiv 0$ as $\ep \to 0$. This is the so-called \emph{double-body flow}, corresponding to the fluid speed where the free surface has been replaced by a rigid plate. 
\end{remark}

\begin{remark} \label{rem:ac}
Note that the leading-order speed \emph{e.g.} in \eqref{eq:q0step} and \eqref{eq:q0tri} contains singularities in the analytic continuation, where $\xi \in\mathbb{C}$. These are associated with the geometrical singularities of the geometry; however, they are not those geometrical singularities themselves. Indeed they correspond to solving \eqref{eq:govset} where the streamfunction has been evaluated on the free surface, $\psi = 0$ or $\xi \geq 0$. As an example, the physical corner of the step corresponding to \eqref{eq:q0step} is at $w = \phi + \im\psi = 
\log b - \im\pi$ but the respective analytically-extended singularity is at $\phi = \log(-b)$. 
\end{remark}
\noindent Additional discussion of the second remark can be found in \S{2} of \cite{trinhNewGravityCapillary2013a}. 

\begin{remark} \label{rem:ac_notation}
Following \cite{chapmanExponentialAsymptoticsGravity2006}, when studying the analytic continuation of the free surface, where $\xi, \phi \in \mathbb{C}$, it is typical to re-annotate $\xi \mapsto \zeta \in\mathbb{C}$ and $\phi \mapsto w \in\mathbb{C}$.
\end{remark}


According to the theory of exponential asymptotics, in the limit $\ep\to 0$, the formation of free-surface waves emerges as a consequence of the Stokes phenomenon, which occurs across so-called Stokes lines; such Stokes lines emerge from critical points or singularities in the analytic continuation of the free surface (cf. the singularities referenced in \cref{rem:ac}). We shall explain these concepts more carefully in the next section. In their work, \cite{chapmanExponentialAsymptoticsGravity2006} consider the local behaviour of the leading-order speed, $q \sim q_0 = q_s$ in the complex plane. Then for stagnation points and flows around corners, locally in the vicinity of the singularity, $(w - w_0) = \text{const.} \times (z - z_0)^k$ for some $k \neq 1$. For a stagnation point, $k = 2$. For flow around a corner, $k = \pi/\beta$ where $\beta$ is the in-fluid angle of the corner. From \cref{rem:q0} and \cref{rem:ac}, the corresponding singularity in the analytic continuation of the free boundary is 
\begin{equation}
	\dd{w}{z} = q_0 \e^{-\im \theta_0} = q_0 = c(w - w_0)^\alpha,
\end{equation}
where $\alpha = (k-1)/k$ and $c$ is constant. 

At this point, \cite{chapmanExponentialAsymptoticsGravity2006} utilise a function, $\chi$, known as the \textit{singulant}, developed via the exponential asymptotics. As $w \to w_0$, the singulant behaves as 
\begin{equation} \label{eq:earlysing}
	\chi \sim -\left[\frac{\im}{c^3(1 - 3\alpha)}\right] (w - w_0)^{1 - 3\alpha}.
\end{equation}
However, as part of the exponential asymptotics procedure, they write: 
\begin{quotation}
\noindent {\itshape Remembering that we only need $\chi$ to vanish at $w = w_0$, we see that we require $\alpha < 1/3$. Only those singularities with $\alpha < 1/3$ will generate Stokes lines and their exponentially small correction terms. [For the case of corner flows] Stokes lines (and gravity waves) are not generated by stagnation points, and will only be generated by corners with in-the-fluid angles greater than $2\pi/3$ [\ldots]}
\end{quotation}

\begin{figure}[htb]
\centering
\includegraphics[scale=1]{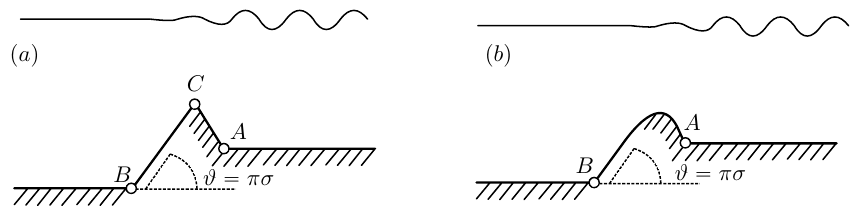}

\caption{Schematics of two-dimensional potential flow over (a) a sharp bump, and (b) a smooth bump in the physical $(x,y)$-plane. Both bumps separate from the horizontal fluid bottom at internal angles of $\vartheta = \pi\sigma$.\label{fig:geoms}}
\end{figure}

Consider now the two scenarios shown in \cref{fig:geoms}. The points A and B, shown in both subfigures, are stagnation points and have no associated Stokes line contribution. Then in the flow past the triangular obstruction in \cref{fig:geoms}(a), downstream waves are expected due to the corner point C. However, in \cref{fig:geoms}(b), there are no apparent (leading-order) exponentially-small waves since there are no further `critical' singularities. According to the above criterion by \cite{chapmanExponentialAsymptoticsGravity2006} there is no obvious mechanism for gravity-wave generation.

The apparent contradiction of the \cite{chapmanExponentialAsymptoticsGravity2006} criterion in connection with the bump geometry was recently raised in the numerical work of \cite{pethiyagodaEfficientComputationTwodimensional2018}. There, it was confirmed that, predictably, surface waves are generated by flows over such topographies. The authors further established the exponential smallness of the downstream waves as $\ep \to 0$; through careful numerical fitting, it was verified that the wave amplitude likely scales as $\sim \mathcal{A} \ep^{\mu/2} \, \e^{-\nu/\ep}$ for constants $\mathcal{A}$, $\mu$, and $\nu$. No further explanation of the underlying mechanism was given, but it was written that (p.~618):
\begin{quotation}
	\noindent \emph{``These results suggest that there is a Stokes line intersecting the free surface even though no Stokes line originates from the two corners in this configuration. As such, there must be a Stokes line structure that is not described by \cite{chapmanExponentialAsymptoticsGravity2006} or elsewhere, leaving an interesting open problem for further research in the area of exponential asymptotics."}
\end{quotation}


Before going further, it would be useful to make a general remark regarding the apparent failing of the criterion presented above. As we interpret it, the issue relates to the ill-posedness of analytic continuation: the smoothed geometry presented in \cref{fig:geoms}(b) seems to be only a slight deformation from the triangular obstruction in \cref{fig:geoms}(a). However, such geometries can introduce quite unpredictable behaviours in the analytic continuation of the associated functions (the $q_s$ in this case). Indeed the smooth geometry, presented later, contains an essential singularity as the velocity potential, $|\phi| \to \infty$, in certain sectors of the complex plane; this is beyond the class of singularities considered in \cite{chapmanExponentialAsymptoticsGravity2006}. Thus in the case of the smooth geometry, waves are produced by a somewhat unexpected singularity (and its associated Stokes line) in the analytic continuation at infinity.



\section{Exponential asymptotics and its consequences}

Above, we have not provided any explanation of exponential asymptotics and the construction of the singulant function, $\chi$, in \eqref{eq:earlysing}. We now provide a short introduction to the use of exponential asymptotics in gravity free-surface flows. 

When $\epsilon = 0$, $\theta = 0$ and indeed this corresponds to the so-called {\itshape double-body flow} where the free surface has been replaced by a rigid plate. In this case, the leading-order speed is given by $q \sim q_0 = q_s$. As was originally noted by \cite{ogilvieWaveResistanceLow1968}, this initial approximation is entirely wave-free. Moreover, as a consequence of the asymptotic procedure, all higher-order algebraic corrections in $\ep$ are also wave-free (this fact Ogilvie and others would later refer to as the ``low-speed paradox"). The resultant water waves, in fact, are expected to be exponentially-small in $\ep$; their amplitudes scale as $\e^{-\text{const.}/\ep}$ in the far field as $\ep \to 0$. This exponential smallness was known by a number of people, Ogilvie included, but the proper techniques for the derivation of such waves was developed in parallel with the later developments of exponential asymptotics as an proper subject of interest. Indeed, in the seminal work of \citet{vb_1977} and \citet{vb_1978}, the authors used techniques in series acceleration and the Shanks transform to derive the exponentially-small contributions. Like many early attempts in beyond-all-orders asymptotics, the prior analyses were correct in spirit, but were not able to derive all features of the waves. 

The proper procedure was presented by  \cite{chapmanExponentialAsymptoticsGravity2006}, who established an exponential asymptotics method for the derivation of the waves in low-Froude problems for two-dimensional free-surface flows. There, a connection is made between the exponentially-small waves and the underlying divergence of the naive asymptotic expansion. It is explained that in the limit $\ep \to 0$, the free-surface waves must arise as a consequence of the Stokes phenomenon---in essence, there exist critical Stokes lines (curves) in the complex plane across which the exponentially-small waves switch on from one side to the other. We now briefly explain these features.

\emph{Divergence and the singulant.} First, it should be remarked that the example geometries given in \eqref{eq:q0step} and \eqref{eq:q0tri} contain singularities in the analytic continuation of $\xi$ or $\phi \in \mathbb{C}$. Consider a standard asymptotic expansion for the unknowns, say $q = q_0 + \ep q_1 + \ep^2 q_2 + \ldots$ and $\theta = \theta_0 + \ep \theta_1 + \ep^2 \theta_2 + \ldots$. By Bernoulli's equation, each subsequent asymptotic order of the perturbative expansion relies upon differentiation of the previous order. Then if $q \sim q_0 = q_s$ has singularities in the complex plane, such singularities tend to increase in strength as more terms of the expansion are taken. Generically, this tends to induce factorial-over-power divergence of the late terms, $q_n$ as $n \to \infty$ \citep{chapmanExponentialAsymptoticsStokes1998}. Inductively argued, as $n \to \infty$, it is posited that
 \begin{align}\label{eqn:bump-ansatz} 
q_n \sim \frac{Q(\zeta)\Gamma(n+\gamma)}{[\chi(\zeta)]^{n+\gamma}} \qquad \text{and} \qquad \theta_n(\zeta) \sim \frac{\Theta(\zeta)\Gamma(n+\gamma)}{[\chi(\zeta)]^{n+\gamma}},
\end{align}
where $\gamma$ is constant, and $Q$, $\Theta$, and $\chi$ are complex-valued functions to be derived. The function $\chi$ is often called the \emph{singulant}, and by assumption, $\chi = 0$ at those singularities, $\zeta, w \in\mathbb{C}$ that induce the divergence of the asymptotic expansion. For instance, from \eqref{eq:q0tri} corresponding to a triangular obstruction we might expect the divergence of the late terms to be driven by the three critical points at $\zeta = -c$, $\zeta = -1$, or $\zeta = -b$. 

\emph{Optimal truncation the exponentially-small terms.} Next, in order to examine the exponentially-small terms, one then optimally truncates the divergent asymptotic expansion and examines the remainder. Thus, for instance, we write $q = q_0 + \ep q_1 + \ep^2 q_2 + \ldots + \ep^{N-1} q_{N-1} + R_N$. In the limit $\ep \to 0$, if $N \to \infty$ is chosen optimally, then it can be verified that the remainder is indeed exponentially small. This procedure of optimal truncation and examination of the remainder is known as {\itshape Stokes line smoothing} \citep{berryUniformAsymptoticSmoothing1989}. Analysis of the remainders thus yields the leading-order exponentially-small terms as:
\begin{equation}\label{eqn:bump-exp-terms}
q_\text{exp} \sim \frac{2\pi\im}{\ep^\gamma}Q(\zeta) \exp\left[ -\frac{\chi(\zeta)}{\ep} \right] \qquad \text{and} \qquad \theta_\text{exp} \sim \frac{2\pi\im}{\ep^\gamma}\Theta(\zeta) \exp\left[ -\frac{\chi(\zeta)}{\ep} \right].
\end{equation}
When evaluated on the physical free surface, the above contribution is supplemented by its complex conjugate, hence yielding a real-valued oscillation. The fact that there is a link between the divergence \eqref{eqn:bump-ansatz} and the exponentials \eqref{eqn:bump-exp-terms} is seemingly unusual, but is a commonality of exponential asymptotics. 

\emph{The criteria for Stokes lines.} Now we may finally explain the key criterion developed in \cite{chapmanExponentialAsymptoticsGravity2006} regarding the existence of Stokes lines. As noted previously, the contributions \eqref{eqn:bump-exp-terms} do not exist everywhere in the complex plane, but are switched-on across Stokes lines (curves). Such Stokes lines are given by those points $\zeta\in\mathbb{C}$ where
\begin{equation} \label{eq:dinglecond}
\Im\chi(\zeta) = 0 \quad \text{and} \quad \Re \chi(\zeta) \geq 0,
\end{equation}
a criterion known as \emph{Dingle's condition} \citep{dingleAsymptoticExpansionsTheir1973}. 

\citet{chapmanExponentialAsymptoticsGravity2006} demonstrate that $\de{\chi}/\de{w} = -\im/q_s^3$. Therefore study of Stokes lines reduces to integrating this relationship from each singularity of interest, and examining the locus satisfying \eqref{eq:dinglecond}. This then leads to our presentation of \eqref{eq:earlysing}. Thus we can see that near the critical point, say $w = w_0$, $\chi \sim \text{const.} (w - w_0)^{1-3\alpha}$. Therefore, in order for a Stokes line to emerge $w = w_0$, it must be the case that 
\begin{equation} \label{eq:criterion}
1 - 3 \alpha > 0. 
\end{equation}
The condition essentially ensures that $\chi \to 0$ as $w \to w_0$ and that the Stokes line emerges on the same Riemann sheet as the physical free surface. Physically, this corresponds to a Stokes line being generated by flows past obstructions with an in-fluid angle of $2\pi/3$. This is the key criterion, referenced by \cite{chapmanExponentialAsymptoticsGravity2006}, that motivates our investigation.

\subsection{On general smooth bodies}

The above criterion \eqref{eq:criterion} seems to suggest that only flows past obstructions with geometrical corners (corresponding to algebraic branch points in the potential) produce waves. 

\begin{figure}[h]
\centering
\includegraphics[width=0.5\textwidth]{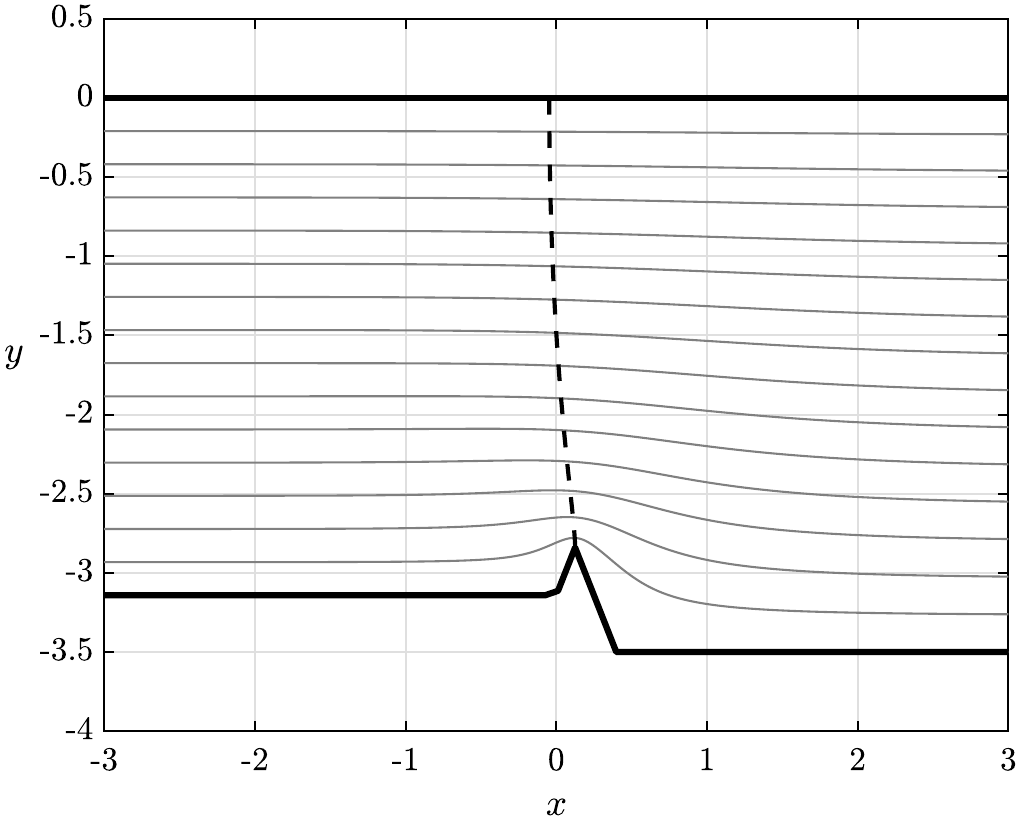}
\caption{Streamlines (thin) for potential flow, with $\epsilon = 0$, over a sharp bump (thick) in the $(x,y)$-plane corresponding to the geometry \eqref{eqn:sharp-qs} with $a = 1$, $b = 3$, and $\sigma = 3/8$. A Stokes line (dashed) emerges from the apex of the bump. Exponentially-small surface waves are expected to follow downstream from the Stokes line. \label{fig:streamline-replace01}}
\end{figure}

Let us consider the following thought experiment. Say we have a geometry  containing a sharp corner which we know to exhibit the Stokes phenomenon and where we expect the prior exponential asymptotics theory to apply. We illustrate such a geometry in \cref{fig:streamline-replace01}. In this example, a Stokes line emanates from the apex of the bump and induces Stokes switching which produces physical waves in the solution. 

\begin{figure}[h]
\centering
\includegraphics[width=0.5\textwidth]{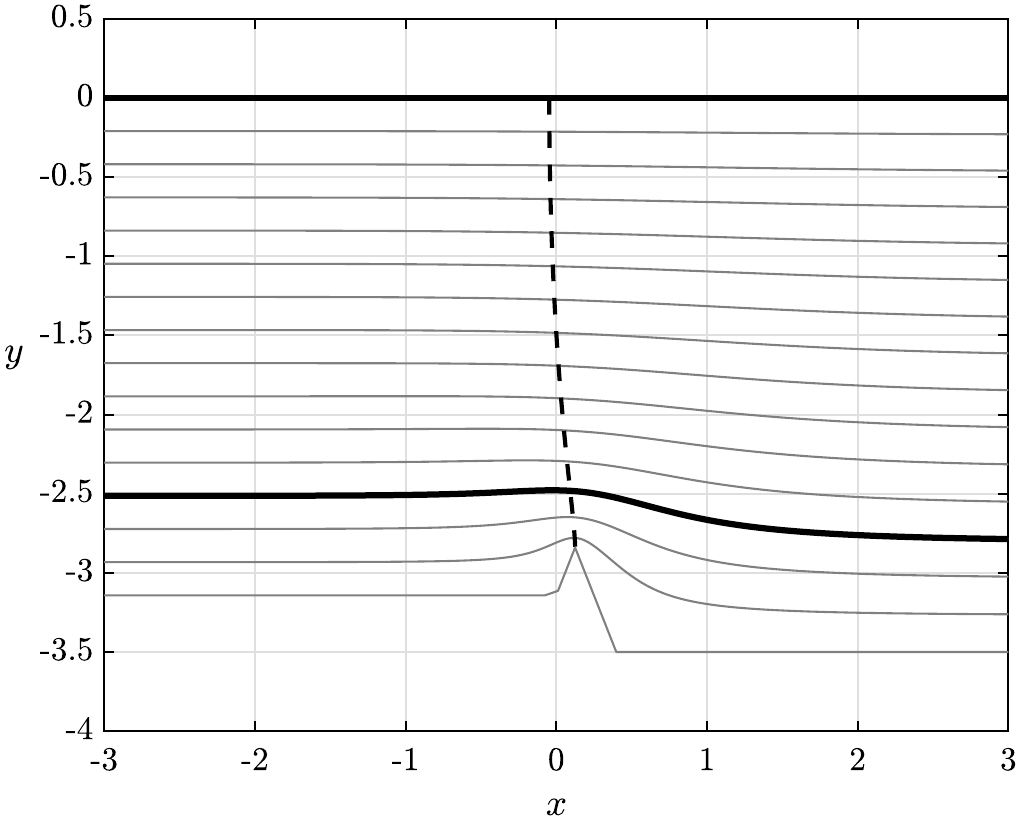}
\caption{Same as in \cref{fig:streamline-replace01} but we consider an equivalent problem in which an in-fluid streamline is treated as a new fluid bottom (shown thick). Despite the fluid bottom now being smooth and locally infinitely differentiable in the physical domain, a Stokes line is nonetheless generated as if from a singular point beneath the fluid bottom. \label{fig:streamline-replace02}}
\end{figure}

Now let us choose a streamline of this flow and consider it the fluid bottom of a new \emph{smooth} geometry. This is shown in \cref{fig:streamline-replace02}. By the properties of potential flow, we expect that the free surface and fluid interior are identical to that of the original problem. However, if we forgot about the original problem it would appear as though Stokes switching had been induced by a completely smooth geometry. The Stokes line must then be interpreted as having been generated from some `imaginary corner' located beneath the physical bottom surface.

This thought experiment illustrates two key points. First, the prior \citet{chapmanExponentialAsymptoticsGravity2006} theory is not limited to obstructions with corners or other `strong' geometrical singularities. Indeed, for every cornered obstruction that produces surface waves, we can use the interchangeable property of the streamlines to find families of fully smooth obstructions which produce those same surface waves and are thus explained by the same Stokes Phenomenon. This is the forwards procedure. 

The second issue the example highlights is regarding the inverse procedure: if we had no knowledge of the original problem (of the cornered geometry), how would we have been able to predict the Stokes switching in the case of flow past a smooth streamline? This issue is related to ill-posedness in analytic continuation, and is further highlighted in the next example. 
\subsection{The mystery of the smoothed bump}

We return to discuss the simple example, sketched in \cref{fig:geoms}, which seems to produce waves for which the \cite{chapmanExponentialAsymptoticsGravity2006} theory does not capture.

Consider the two geometries illustrated in \cref{fig:geoms}. \Cref{fig:geoms}(a) is the triangular obstruction protruding from the bottom of the channel, with the first stagnation point associated with an angle of $\pi\sigma$. In \cref{fig:geoms}(b), we consider a so-called \emph{`smooth bump'}, where the apex of the bump is now rounded. More precisely, the sharp bump is produced by specifying that the angle of the topography is piecewise constant and discontinuous at the corners, A, B, and C; the smooth bump is produced by specifying continuous angles except at A and B (where no Stokes lines are generated). See \cref{sec:2dbump_formulation} for precise details of the geometry.

\begin{figure}[h]
\centering 
\includegraphics[scale=1]{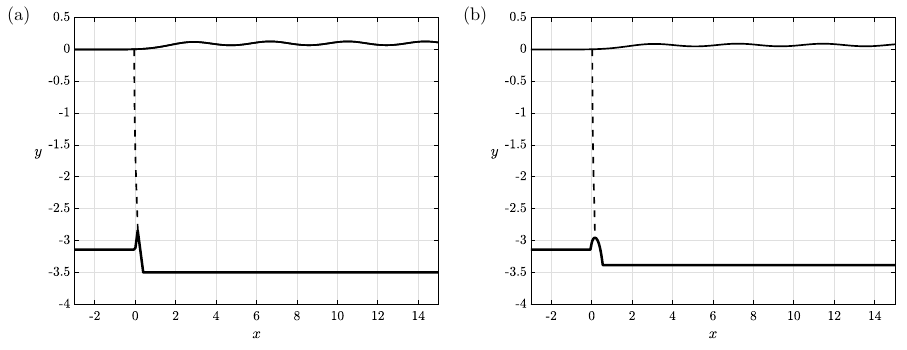}
\caption{Numerical free-surface solutions for flow over a sharp bump (a) and smooth bump (b). Solutions are computed using the boundary-integral numerical procedure discussed in \cite{yyanis_thesis}. Solutions, corresponding to geometries in \cref{sec:2dbump_formulation}, are calculated for parameters $\ep = 0.8, \sigma = 3/8, a = 1, b = 3$ using a finite difference scheme. The (projected) Stokes lines, as calculated from the methodology presented in this work, are shown dashed. \label{fig:bump-num}}
\end{figure}

Numerical solutions of the full water-wave equations \eqref{eq:govset} are presented in \cref{fig:bump-num}, where we confirm the predictable presence downstream waves for flows past both obstructions. These solutions are computed for the Froude parameter of $\ep = 0.8$ and geometrical parameter $\sigma = 3/8$. Indeed, based on these snapshots, there is no reason to believe that the geometry of a smooth bump is exceptional. The numerical computations presented later in this work confirm the exponential smallness of the oscillations in the limit $\ep \to 0$. 

The criterion \eqref{eq:criterion} indicates that, in order for a Stokes line to extend from a critical point in the geometry into the fluid, the local in-fluid angle must exceed $2\pi/3$. Thus, for the case of the sharp bump, so long as $\sigma > 1/3$, no Stokes line is expected from the points A and B. In such cases, the leading-order exponential-small free-surface waves are wholly attributed to the sharp corner on the apex of the bump, C, where the in-fluid angle is greater than $2\pi/3$ [cf. \cref{fig:bump-num}(a)]. 

However, in the case of the smooth bump shown in \cref{fig:bump-num}(b), both local in-fluid angles at A and B are $5\pi/8$, and consequently, not expected to generate Stokes lines. Despite there being no further obvious critical points in the geometry, exponentially small waves are nonetheless present downstream. In this case, to what can we attribute the Stokes switching, and is there some `location' or a particular feature of the geometry which induces the Stokes line? As we shall show later in this work, there does exist a Stokes line for the smooth bump that, when projected into physical space, emerges from the geometry and intersects the free surface (as shown in \cref{fig:bump-num}). However, like the thought experiment presented in the previous section, the origin of this Stokes line is entirely unphysical.


%

\section{Mathematical formulation} \label{sec:2dbump_formulation}

For completeness, we summarise the essential mathematical formulation of steady, irrotational, incompressible, free-surface gravity flow over a geometrical fluid bottom. We shall consider the case of flow over a triangular bump (\cref{fig:geoms}a), and a rounded bump (\cref{fig:geoms}b). After suitable nondimensionalisation, the governing equations are formulated in terms of the velocity potential, $\phi$,
\begin{subequations}\label{whole-problem}
\begin{align}
\nabla^2 \phi = 0 
& \qquad \text{in the fluid,} \label{laplace}
\intertext{with the kinematic and dynamic (Bernoulli) conditions on the free-surface, $z=\eta(x,y)$,} 
\nabla \phi \cdot \bm{n} = 0  & \qquad \text{on $z = \eta(x,y)$ \text{and the fluid bottom}} \label{kinematic}\\
  \frac{\epsilon}{2} (|\nabla \phi|^2-1) + z = 0 
  &\qquad \text{on $z = \eta(x,y)$}, \label{bernoulli}
\end{align}
\end{subequations}
where $\bm{n}$ is the unit outward-pointing normal to the surface. The nondimensional parameter, $\epsilon$, denotes the square of the Froude number, and measures the relative balance between inertial and gravitational forces (which are assumed to act in the negative $z$ direction),
\begin{equation}
\epsilon = \frac{U^2}{gL}.
\end{equation}
As such, the low-Froude limit corresponds to $\epsilon \to 0$.

We introduce the physical variables in complex form, $z = x+\im y$, and the complex potential $w = \phi +\im\psi$, where $\psi$ is the streamfunction. In this manner $\psi = 0$ and $\psi = -\pi$ define the free-surface and the fluid bottom respectively, while $-\pi<\psi < 0$ defines the fluid region. Complex velocity is given by 
\begin{equation}\label{eqn:2d-dwdz-bump}
\dd{w}{z} = u -\im v = qe^{-\im\theta},
\end{equation}
\noindent The governing equations for potential free-surface flow in a channel can be found in \cite{chapmanExponentialAsymptoticsGravity2006}. Recall that solutions are specified in terms of the streamline speed, $q$, and angle, $\theta$. In non-dimensional form, the governing equations, applied on the free surface, have been given previously in \eqref{eq:govset}.

\begin{figure}
\centering
\includegraphics[scale=1]{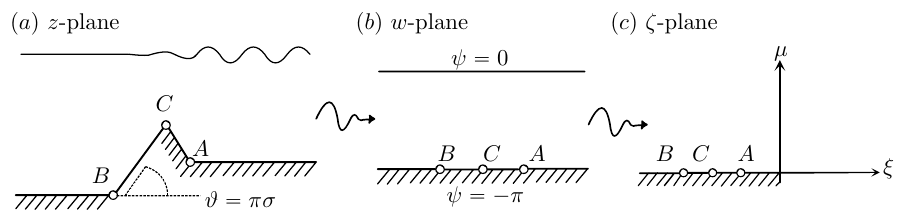}





\caption{Flow over a bump represented in the physical $z = (x + \im y)$-plane (a), potential $w = \phi + \im\psi$-plane $(b)$, and the upper-half $\zeta$-planes $(c)$. The physical region is mapped to potential plane through the a-priori unknown map $w^\prime(z)= q\e^{-\im \theta}$, and then to the upper half $\zeta$-plane using the conformal map $\zeta = \e^{-w}$.\label{fig:sharp-maps}}
\end{figure}

The free-surface and channel bottom are horizontal axes in potential space, defined by fixed values of the streamfunction, $\psi = 0$ and $\psi = -\pi$ respectively. In addition, the flow in the complex potential $w$-plane is transformed to the upper half-$\zeta$-plane plane via the map $\zeta = \e^{-w}$ where $\zeta = \xi + \im\mu$. Under this transformation, the set of equations \eqref{eq:govset} are applied on $\psi = 0$ and $-\infty < \phi < 0$ or equivalently $\mu = 0$ and $0 < \xi < \infty$. A visualisation of the associated conformal maps is given in \cref{fig:sharp-maps}.

\subsection{The shape functions}\label{sect:shape-functions-bump}

\noindent We recall from the previous discussion that the \emph{shape function}, $q_s$, for a particular geometry is defined via \eqref{govset_int}. We now define the two geometries in \cref{fig:geoms} by specification of the streamline angle and derive the respective shape functions. 

\subsubsection{Flow over a sharp bump}\label{sect:sharp}

\begin{figure}[h]
\centering
\includegraphics{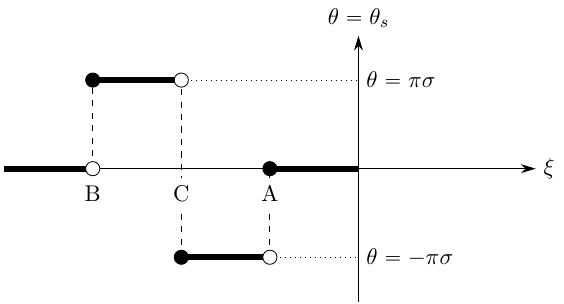}
\caption{The fluid bottom streamline function $\theta_s = \theta_{\text{sharp}}$ which specifies the geometry of the sharp bump. The function contains discontinuities at $\xi = -b$ (point B), $\xi = -c$ (point C), and $\xi = -a$ (point A). \label{fig:thetas_sharp}}
\end{figure}

\noindent The geometry of the sharp bump is specified using two constant streamline angles between the three critical points on the fluid bottom, $\zeta = \xi = \{-b, -c, -a\}$ where $0 < a < c < b$. These three points correspond to the respective points B, C, and A shown in \cref{fig:geoms}a. We note that we have the freedom to choose the origin of the potential plane, and hence we select the point $(\phi, \psi) = (0, -\pi)$ to correspond to the first stagnation point, A. Consequently, $a = 1$ without loss of generality. If we define the initial slope of the bump by the streamline angle $\sigma\pi$, then then we may write the streamline angle on the fluid bottom as
\begin{equation}\label{eqn:sharp-theta}
\theta =
\begin{cases}
0 & \text{if } \xi\in(-\infty,-b)\cup(-a,0),\\
\pi\sigma & \text{if } \xi\in[-b,-c],\\
-\pi\sigma & \text{if } \xi\in[-c,-b],
\end{cases}
\end{equation}
with $0 < a < c < b$. The function is illustrated in \cref{fig:thetas_sharp}. 

Using this, we compute the shape function, $q_s$, from \eqref{govset_int}, which yields
\begin{equation}\label{eqn:sharp-qs}
q_s = \left[\frac{(\xi+a)(\xi+b)}{(\xi+c)^2}\right]^{\sigma}.
\end{equation}
%

In this work, we also choose
\begin{equation} \label{eq:cdef}
    c = \frac{a + b}{2},
\end{equation}
in order to reduce the parameter space and provide a closer relationship to the smooth bump presented next. As noted, $a = 1$ above by choice of scaling.


In this work, we shall focus on the case of $\sigma = 3/8$. This is chosen as a convenient value such that $1/3< \sigma < 1/2$. The lower limit ensures that the bottom corners $B$ and $A$ do not admit an internal fluid angle greater than $2\pi/3$ and therefore do not generate Stokes lines in the fluid, while the upper limit is a geometrical requirement of the sharp bump.

Finally, note that from the $(q, \theta)$ boundary-integral formulation, we may recover flows within physical $(x, y)$-space by integrating the complex velocity relationship \eqref{eqn:2d-dwdz-bump}.

\subsubsection{Flow over a smooth bump}\label{sect:smoothed}

\begin{figure}[h]
\centering
\includegraphics{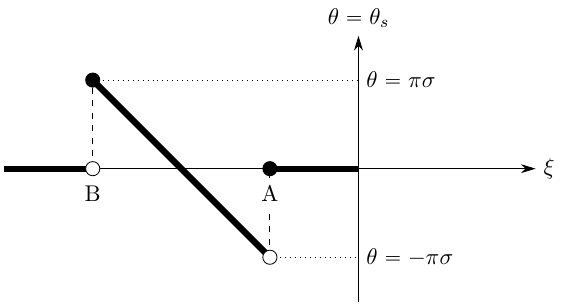}
\caption{The fluid bottom streamline function,   $\theta = \theta_s$, which specifies the geometry of the smooth bump. The function contains discontinuities at $\xi = -b$ (point B) and $\xi = -a$ (point A). \label{fig:thetas_smooth}}
\end{figure}

\noindent The formulation for the geometry of the smooth bump is proposed in an analogous manner, and the angles are illustrated in \cref{fig:thetas_smooth}.

In this case, we model the geometry with two points, $\xi= \{-b, -a\}$ corresponding to the physical stagnation points B and A shown in \cref{fig:geoms}. We assume that the first stagnation point at B is characterised by streamline angle $\theta = \sigma\pi$. In this case, the functional form of $\theta$ along the boundary is
\begin{equation}\label{eqn:smoothed-theta}
\theta =
\begin{cases}
0 & \text{if } \xi\in(-\infty,-b)\cup(-a,0),\\
\sigma\pi-\frac{2\sigma\pi}{b-a}(\xi+b) & \text{if } \xi\in[-b,-a].
\end{cases}
\end{equation}
Using this, we compute the shape function, $q_s$, yielding
%
\begin{equation}\label{eqn:smoothed-qs}
q_s(\xi) = A\left(\frac{\xi+a}{\xi+b}\right)^{B(\xi+c)},
\end{equation}
where $A = \e^{2\sigma}$, $B = 2\sigma/(b-a)$, and $c = (a+b)/2$ as in \eqref{eq:cdef}. As with the sharp bump, we are free to scale the problem and select $a = 1$ without loss of generality. 

Note that specification in terms of potential functions means the geometry will not be perfectly semi-circular. However $q_s$ characterises the qualitative features we are interested in: namely a smooth bump separating-from and returning-to a flat bottom. The initial streamline angle of separateion at B is $\sigma\pi$. We choose the same value of $\sigma = 3/8$ as used in the sharp bump to allow for direct comparison between the two problems.

\section{The development of a simplified model}\label{sect:exp-asy-review}

In this work, we leverage a powerful simplification that was presented in \cite{trinhReducedModelsGravity2017}, and that has been used in a number of wave-structure interaction approaches with exponential asymptotics, \emph{e.g.} \cite{trinhTopologicalStudyGravity2016,jamshidiGravityCapillaryWaves2020}.

The key idea is as follows. In order to study the behaviour of the asymptotic expansions in the complex plane, the boundary-integral equation \eqref{govset_int} must first be written as 
\begin{equation}\label{eqn:bump-governing-bdint_AC}
\log\left(q\right) - \im \theta=\log q_s -\hat{\mathcal{H}}[\theta],
\end{equation}
where $\hat{\mathcal{H}}$ denotes the analytic continuation of the Hilbert transform, given by 
\begin{equation}\label{eqn:bump-hilbert-ac}
\hat{\mathcal{H}}[\theta](\zeta)=\frac{1}{\pi}\int_{0}^{\infty}\frac{\theta(\xi^\prime)}{\xi^\prime-\zeta}\de \xi^\prime.
\end{equation}
It can be verified that the above formulae reduces to \eqref{govset_int} for $\zeta = \xi \in \mathbb{R}$ when the real axis is approached from the upper half-plane.

During the exponential asymtotics approach by \cite{chapmanExponentialAsymptoticsGravity2006}, it is explained that as $n \to \infty$, determination of the late terms \eqref{eqn:bump-ansatz} does not actually require the complex Hilbert transform, $\hat{\mathcal{H}}$ (to the relevant orders) since it is exponentially subdominant, \emph{e.g.} $\hat{\mathcal{H}}\theta_n$ is subdominant to $\theta_n$ for $n \to \infty$. Consequently, the functional forms of the crucial terms in the ansatz \eqref{eqn:bump-ansatz} and exponential remainders \eqref{eqn:bump-exp-terms}, namely $\chi$, $Q,$ and $\Theta$, can be derived independently of the Hilbert term. What this suggests is that it may be possible to develop simpler models of the exponential asymptotics by ignoring the Hilbert term, $\hat{\mathcal{H}}$, in \eqref{eqn:bump-governing-bdint_AC} and using 
\begin{equation} \label{eqn:bump-toysub}
\log\left(q\right) - \im \theta \sim \log q_s.
\end{equation}
%
%
%
and then substituting the above into Bernoulli's equation. This essentially results in simpler complex-plane models that, in an asymptotically justifiable sense, approximate the precise details of the $\ep \to 0$ limit and provide accurate approximations to the full water-wave equations at low Froude numbers.





Following \cite{trinhReducedModelsGravity2017}, we review the derivation of these models. 

In the $\ep\to 0$ limit, we expect the solution to split into a base series, representing the wave-free solution, and a remainder term representing exponentially small gravity waves. The base series is the standard asymptotic series truncated at order $\ep^{N-1}$, while the remainders are $\Oh(\ep^N)$,  
\begin{align}\label{eqn:bump-truncated}
q =\underbrace{\sum_{n=0}^{N-1}\ep^n q_n}_{q_r}+\overline{q} \qquad \text{and}\qquad \theta = \underbrace{\sum_{n=0}^{N-1}\ep^n \theta_n}_{\theta_r}+\overline{\theta}.
\end{align}
 We wish to obtain an integral representation of the remainder terms, $\overline{q}$ and $\overline{\theta}$. Substituting the truncated forms \eqref{eqn:bump-truncated} into the governing Bernoulli equation \eqref{eq:mybern} and boundary-integral equation \eqref{eqn:bump-governing-bdint_AC} produces
 \begin{subequations}
 \begin{equation}
\epsilon\left(q_{r}^{2} q_{r}^{\prime}+2 q_{r} q_{r}^{\prime} \bar{q}+q_{r}^{2} \bar{q}^{\prime}\right)+\sin \theta_{r}+\bar{\theta} \cos \theta_{r}=\Oh\left(\bar{q}^{2}, \bar{\theta}^{2}\right), 
 \end{equation}
 \begin{equation}
 \log q_{r}-\log q_{s}+\frac{\bar{q}}{q_{r}}+\hat{\mathcal{H}}\left[\theta_{r}\right]+\hat{\mathcal{H}}[\bar{\theta}]-\mathrm{i}\left(\theta_{r}+\bar{\theta}\right)=\Oh\left(\bar{q}^{2}\right),
 \end{equation}
 \end{subequations}
 where primes ($'$) correspond to derivatives in $w$. The two above equations are then combined to give a single integro-differential equation:
 \begin{equation}
\epsilon\left(q_{r}^{\prime}+2\frac{q_{r}^{\prime} \bar{q}}{q_{r}}+\bar{q}\right)+\frac{\sin \theta_{r}}{q_{r}^{2}}
 -\frac{\cos \theta_{r}}{q_{r}^{2}}\left(\mathrm{i} \log \frac{q_{r}}{q_{s}}+\mathrm{i} \frac{\bar{q}}{q_{r}}+\theta_{r}+\mathrm{i} \hat{\mathcal{H}}\left[\theta_{r}\right]+\mathrm{i} \hat{\mathcal{H}}[\bar{\theta}]\right) = 0.
\end{equation}

We now expand in powers of $\ep$ using $q_r = q_s+\ep q_1+ \ep^2 q_2 + \cdots$ and $\theta_r  = 0 +\ep \theta_1+ \ep^2 \theta_2 + \cdots$. This yields an ODE for $\overline{q}$ with a forcing term dependent on $\overline{\theta}$:
\begin{equation}\label{eqn:trinh2017-reduced}
\epsilon \bar{q}^{\prime}+\left[-\frac{\mathrm{i}}{q_{s}^{3}}+\epsilon\left(2 \frac{q_{s}^{\prime}}{q_{s}}+3 \mathrm{i} \frac{q_{1}}{q_{s}^{4}}\right)+ \Oh\left(\epsilon^{2}\right)\right] \bar{q}=\mathcal{R}(w ; \hat{\mathcal{H}}[\bar{\theta}]) + \Oh\left(\bar{\theta}^{2}, \bar{q}^{2}\right).
\end{equation}
In the above, the forcing term $\mathcal{R}$ is given by
\begin{subequations}\label{eq:RHSall}
\begin{align} 
    \mathcal{R}(w ; \hat{\mathcal{H}}[\bar{\theta}]) &= -\mathcal{E}_{\text{bern}}+\mathcal{E}_{\text{int}}\frac{\cos\theta_r}{q_r} \\
    \mathcal{E}_{\text{bern}} &= \ep q_r^\prime+\frac{\sin\theta_r}{q_r^2},\\
    \mathcal{E}_{\text{int}} &= \log{q_r}-\im\theta_r-(\log q_0-\hat{\mathcal{H}}[\theta_r]).
\end{align}
\end{subequations}
The above equation \eqref{eqn:trinh2017-reduced} is valid for a truncation of in \eqref{eqn:bump-truncated} with $N \geq 2$. For the case of $N = 1$, see \eqref{eqn:N1} below. 

The ODE \eqref{eqn:trinh2017-reduced} may be solved by means of an integrating  factor to give
\begin{equation}\label{eqn:integrating-factor-soln}
    \overline{q}(w) = Q(w)I(w) \e^{-\chi/\ep},
\end{equation}
where
\begin{subequations}\label{eqn:bdint-chiQI}
\begin{align}
\chi(w) &= -\im\int_{w_0}^w\frac{\de \varphi}{q_s^3(\varphi)},\label{eqn:bump-chi}
\\
Q(w) &= \frac{\Lambda}{q_s(w)^2}\exp\left\{-3\im\int_{w_\star}^{w}\frac{q_1(\varphi)}{q_s^4(\varphi)}\de \varphi\right\},\label{eqn:bump-Q}\\
I(w) &= \int_{-\infty}^{w}\mathcal{R}(\varphi;\hat{\mathcal{H}}[\overline{\theta}])\left(\frac{1}{Q(\varphi)}+\Oh(\ep)\right)\e^{\chi(\varphi)/\ep}\de \varphi.\label{eqn:trinh-I}
\end{align}
\end{subequations}
We remind the reader that $q_s = q_0$. The starting point of integration in \eqref{eqn:bump-chi}, $w_0$, is a singular point of $q_0$, while in \eqref{eqn:bump-Q} $w_\star$ is an arbitrary starting point of integration. We note that a different value of $w_\star$ changes the value of $\Lambda$. It is often convenient to take $w_\star = w_0$, but the integral is not always defined at this point. We note that \eqref{eqn:integrating-factor-soln} is not a closed form solution since $Q$ and $I$ depend on the Hilbert transform of the unknown $\overline{\theta}$ (the former due to the presence of $q_1$). Despite this, the integrals \eqref{eqn:bdint-chiQI} show that a crucial part of the solution, namely the \emph{singulant} function, $\chi$, can be resolved without dependence on the Hilbert transform. Indeed this singulant function is the same as derived by \cite{chapmanExponentialAsymptoticsGravity2006}. 

\subsection{Choosing the truncation point N}

At this point, the ODE \eqref{eqn:trinh2017-reduced} and integral forms \eqref{eqn:integrating-factor-soln} are exact. However, as noted, they are implicit due to the unknown $\mathcal{R}$ in \eqref{eq:RHSall}. Now, different truncation choices for $N$ in \eqref{eqn:bump-truncated} produce different models. Once $N$ is chosen, and the reduced $\mathcal{R}$ calculated, then the exponential asymptotic analysis can be done via a steepest descent calculation of \eqref{eqn:integrating-factor-soln}. 

Suppose the leading-order exponentially-small waves are given by $q_\text{exp} \sim \mathcal{A} \mathcal{F}(w) \e^{-\chi/\ep}$ (plus its complex conjugate). In \cite{trinhReducedModelsGravity2017}, it is explained that truncation at $N = 1$ reproduces the singulant behaviour correctly, i.e. $\e^{-\chi/\ep}$,  whereas truncating at $N = 2$ reproduces in addition the functional prefactor, $\mathcal{F}$. Computation of $\mathcal{A}$ requires optimal truncation and hence an argument with $N \to \infty$ as $\ep \to 0$.

We are focused on explaining the dominant wave phenomena in the case of the sharp bump or smoothed bump cases in \cref{fig:geoms}. Although different truncation models will change the functional prefactor of the exponential switching, but much of the analytical `spirit'---notably the singulant analysis and Stokes lines---is preserved with the simplest case of $N = 1$. This will be our focus for the remainder of the paper.

The case of truncation $N = 1$ is most easily derived by setting the $\ep$ terms on the left hand-side of \eqref{eqn:trinh2017-reduced} to zero, and noting that $\mathcal{E}_\text{bern} = \ep q_0'$ and $\mathcal{E}_\text{int} = 0$. Then, written in either $w$ or $\zeta = \e^{-w}$, we have
\begin{equation}\label{eqn:N1}
\ep \overline{q}' + \left[-\frac{\im}{q_0^3}\right] \overline{q} \sim -\ep q_0' \qquad \text{or} \qquad \ep\dd{\overline{q}}{\zeta} + \left[\frac{\im}{\zeta q_0^3}\right]\overline{q} \sim - \ep \dd{q_0}{\zeta},
\end{equation}
where $q_0 = q_s$ is defined by the fluid geometry in question. For the two bump geometries studied in this work, the respective shape functions $q_s$ will be \eqref{eqn:sharp-qs} and \eqref{eqn:smoothed-qs}. By the same fashion as leading to \eqref{eqn:integrating-factor-soln}, we may now write down the explicit solution. It is simplest to do so for integration in the $\zeta$-plane:
  \begin{equation}\label{eqn:mainint}
  \overline{q}(\zeta)\sim-\left[\int_{S}^\zeta \dd{q_0}{\tilde{\zeta}} \, \e^{\chi(\tilde\zeta)/\ep}\de \tilde\zeta\right] \e^{-\chi(\zeta)/\ep},
  \end{equation}
where, from \eqref{eqn:bump-chi},
\begin{equation}\label{eqn:N1chi}
\chi(\zeta) = \int_{\zeta^\star}^{\zeta}\frac{\im}{\tilde\zeta q_0^3(\tilde\zeta )}\de \tilde{\zeta}.
\end{equation}
In regards to the above integral, we keep the following considerations in mind. 
\begin{enumerate}
\item In both integrals \eqref{eqn:mainint} and \eqref{eqn:N1chi}, the dummy variables correspond to integration in the $\zeta$-plane. The initial point of integration in $\chi$ will generally be a critical point in the geometry (at least in the case of the sharp bump, where it is known) (for example the apex, C, of the sharp bump). 

\item The initial point, S, of the integral solution \eqref{eqn:mainint} can be chosen to be some large positive non-degenerate point on the free surface ($\tilde\zeta\geq0$). Later in the analysis, we impose the uniform flow condition in the upstream far-field by taking $S\to\infty$. 

\item Note that on the free surface, the fluid speed $q_0$ is purely real. This implies that  $\Re\chi = \text{const.}$ on the free surface. Consequently integration from the singularity to any point on the free surface should predict the amplitude dependence of the waves on $\ep$. This will provide a verification of $\chi$ later. 

\item In the situation of flow over the smooth bump, the absence of an obvious Stokes-line generating point in the geometry makes the choice of $\zeta^*$ more difficult. As shall be discussed further in \cref{sect:smoothed}, the start point should be chosen as an essential singularity at infinity since it will turn out that this is where the Stokes line originates from.
\end{enumerate}
We now outline the process of asymptotically approximating the integral solution \eqref{eqn:mainint} using the method of steepest descents.

\section{Steepest descent analysis of the $N=1$ truncation model} \label{sec:2dbump_steep1}



In the following sections, we study the asymptotics of \eqref{eqn:mainint} in the limit $\ep \to 0$, where $q_0$ takes the case of the sharp bump \eqref{eqn:sharp-qs} or smooth bump \eqref{eqn:smoothed-qs}. Recalling that $\zeta$ serves as the upper limit of integration in \eqref{eqn:mainint}. We demonstrate that, as $\ep \to 0$, there are certain critical values of $\zeta$ across which the dominant contributions to the solution qualitatively change. 

\begin{figure}[htb]
\centering
\includegraphics[scale=1]{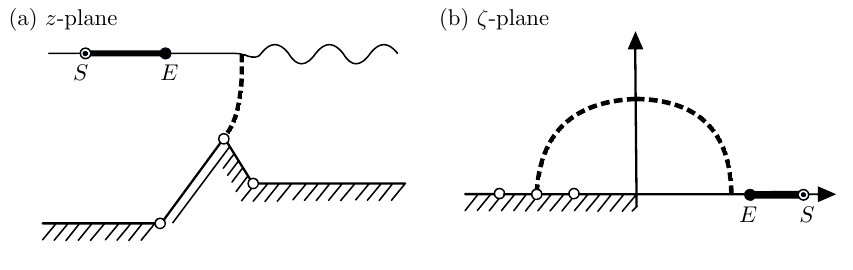}
\caption{Illustration of \emph{Case I} scenario, where the endpoint, $E$ of integration is upstream of the dashed line, as shown in the physical plane (left) or $\zeta$-plane (right). The integration (thick solid) from $S$ to $E$ results in a non-oscillatory solution. The dashed line is a (projection of the) Stokes line.\label{fig:sd-case1}}
\end{figure}

In the first instance, which we refer to as \emph{Case I}, we demonstrate that when $\zeta$ is chosen sufficiently far upstream in the flow (as illustrated in \cref{fig:sd-case1}), the integral has dominant contributions from the endpoints of the integration. This produces the base wave-free solution (\emph{i.e.} the integral approximation process recovers the standard asymptotic approximation).

\begin{figure}[h]
\centering
\includegraphics[scale=1.0]{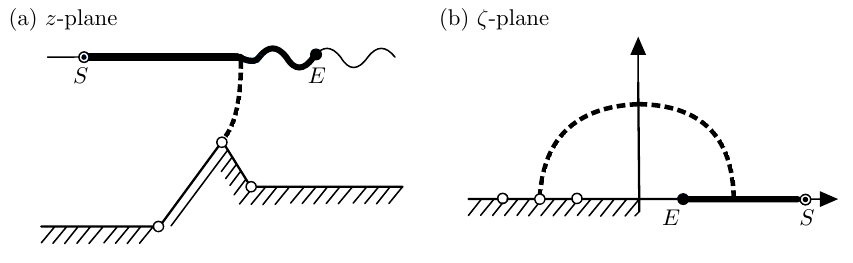}
\caption{Illustration of \emph{Case II} scenario, where the endpoint, $E$ of integration is downstream of the dashed line, as shown in the physical plane (left) or $\zeta$-plane (right). The integration (thick solid) from $S$ to $E$ results in an oscillatory solution. The dashed line is a (projection of the) Stokes line.\label{fig:sd-case2}}
\end{figure}

 In contrast, in the second instance which we refer to as \emph{Case II}, if $\zeta$ is chosen sufficiently downstream in the flow (as illustrated in \cref{fig:sd-case2}), then the integral \eqref{eqn:mainint} has additional contributions from critical other points in $\zeta\in\mathbb{C}$. These contributions are associated with singularities in the leading-order speed, $q_0$, and correspond to picking up exponentially small waves in the solution \eqref{eqn:mainint}. The dividing curve, dashed red in \cref{fig:sd-case1} and \cref{fig:sd-case2} corresponds to a projection of the Stokes line into physical space. 

The following two sections address the problems of flow over the sharp bump (\S\ref{sect:sharp-sd}) and flow over the smooth bump (\S\ref{sect:smooth-sd}). Each section is structured as follows.
\begin{enumerate}
  \item An overview of the Riemann structure of the singulant $\chi$ in the $\zeta$-plane. 
  \item A steepest descent analysis of the \emph{Case I} scanario: where the end point of integration lies at a \emph{downstream} point of the flow. In this case we will demonstrate that the asymptotic approximation to the solution \eqref{eqn:mainint} is of the form
  \[ 
  \ql \sim \text{base solution.}
  \]
  \item A steepest descent analysis of the \emph{Case II} scanario: where the end point of integration lies at an \emph{upstream} point of the flow. In this case we will demonstrate that the asymptotic approximation to the solution \eqref{eqn:mainint} is of the form
  \[ 
  \ql \sim \text{base solution} + \text{exponentially small ripples.} 
  \]
\end{enumerate}
Finally, we verify the asymptotic predictions obtained through the steepest descent analysis by using numerical results for the full boundary-integral system \eqref{eq:govset}. 

\section{Steepest descents for the sharp bump}\label{sect:sharp-sd}

\begin{figure}[htb]
\centering
\includegraphics{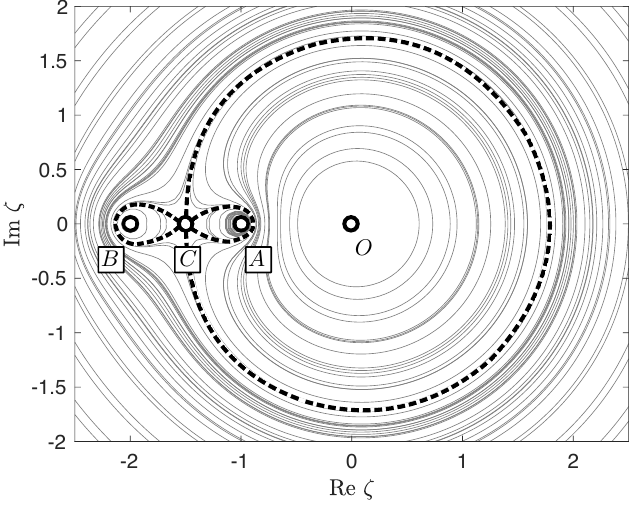}
\caption{A contour plot of $\Im\chi$ in the case of the sharp bump with $\sigma = 3/8$. The bump corners $(A, B, C)$ are located at $\zeta = -1, -2, -3/2$. Here $\chi$ is given by \eqref{eq:q0_sharp_steep}. The free-surface and fluid bottom lie on the positive- and negative-real-axes respectively, while upper-half-plane is associated with the inner-fluid region. Three homoclinics (dashed) emanate from the critical point $C$, corresponding to the apex of the bump. The central homoclinic intersects the free surface and plays a crucial role in Stokes switching. As we demonstrate later, this is the Stokes line associated with the switching-on of downstream ripples. A selection of constant-phase paths, $\Im\chi = \text{const.}$ is indicated by grey contours---these represent steepest descent/ascent paths. \label{fig:sharp-conts}}
\end{figure}

We study \eqref{eqn:mainint} with $q_0$ from 
\begin{equation}\label{eq:q0_sharp_steep}
q_0(\zeta) = \left[\frac{(\zeta+1)(\zeta+b)}{(\zeta+c)^2}\right]^{\sigma}.
\end{equation}
 We will focus on the case in which $b = 2$, $c = (1 + b)/2$, and $\sigma = 3/8$. 
A sample of the level sets of $\Im\chi$ is shown in \cref{fig:sharp-conts}. Contours generally form closed loops with the Riemann structure induced by the critical points at $B, C, A,$ and the origin. Points $A$ and $B$ correspond to the bottom corners of the sharp bump, while the point $C$ corresponds to the bump apex. The logarithmic singularity at the origin is an artifact of the mapping $\zeta = \e^{-w}$ from the potential plane to the $\zeta$-plane and is of little consequence in the following analysis. We now examine the critical points $B, C,$ and $A$, which play an important role in the steepest descents analysis.

\emph{On the singularities.} In \cref{fig:sharp-conts} we choose branch cuts from the point $A$ leftwards and from the point $B$ rightwards, and choose the logarithmic branch cut from the origin to extend diagonally into the fourth quadrant. Let us introduce the index notation to denote the individual Riemann sheets, $\chi_{(k_1, k_2, k_3, k_4)}$. According to this notation the primary sheet $\chi_{(0,0,0,0)}$ is taken as that which contains the physical fluid, and the indexing numbers $k_i\in\mathbb{Z}$ indicate the number of anticlockwise rotations around the branch points $\zeta = Z_i = \{-b, -c, -a, 0\}$ needed to access the particular sheet from $\chi_{(0,0,0,0)}$. There are three homoclinics which emanate from $\zeta = -c$ and are indicated with red dashed lines in \cref{fig:sharp-crossing}. These partition the plane into four regions. 
The local angle of emergence of these contours can be determined by setting $Z = \zeta + c$ in \eqref{eq:q0_sharp_steep} and studying $Z\to0$. 
From \eqref{eqn:N1chi}, $\chi$ behaves like
\begin{equation}
\chi \sim-\frac{\im}{(6\sigma+1) c}(a-c)^{-3\sigma}(b-c)^{-3\sigma}Z^{(6\sigma+1)}\quad\text{as $Z\to0$}.
\end{equation}
Setting $Z=\rho\e^{\im\alpha}$ and imposing $\Im\chi = 0$, we require $\sin\left(-3\pi\sigma + (6\sigma+1)\alpha\right) = 0$, where we have set $\Arg(a-c) = \pi$ according to the choice of vertical branch cuts. This allows us to find the local angles of emergence,
\begin{equation}
\alpha = \frac{(n+3\sigma)\pi}{6\sigma+1}, \quad n = 0, \, 1, \, \dots,5.
\end{equation}
 We note the additional criterion for a Stokes line requiring $\Re\chi>0$, which corresponds to odd values of $n$. 

\emph{On the constant-phase contours.} In \cref{fig:sharp-conts}, the level sets of $\Im\chi$ are classified by their position relative to the three homoclinics from point $C$. First, within the interiors of the rightmost and leftmost homoclinics (emanating from $C$ at local angles $3\pi/13$ and $11\pi/13$ respectively), constant-phase contours are closed loops around $A$ and $B$ respectively. Additionally, there are constant-phase contours located in between the leftmost and central homoclinic, and these form closed paths around the pole at the origin.

Outside the homoclinics, constant-phase paths form closed contours which encirlcle all four critical points. We note that the points $A$ and $B$, which correspond to stagnation points, are not saddle-points of the integrand since $\chi^\prime$ is nonzero here. We now examine two distinct cases in which the integration end point is \emph{upstream} (\emph{Case I}) and \emph{downstream} (\emph{Case II}) of the central homoclinic from $C$. We demonstrate that the inclusion of oscillatory terms in the asymptotic approximation to the solution depends on the proximity of the end point of integration to this curve. From this we may infer that the central homoclinic is a Stokes line which induces oscillatory behaviour in the solution. Moreover, we can verify this by checking that $\Im\chi = 0, \Re\chi>0$ here.

\afterpage{
\begin{figure}[t]
\includegraphics[scale=1.0]{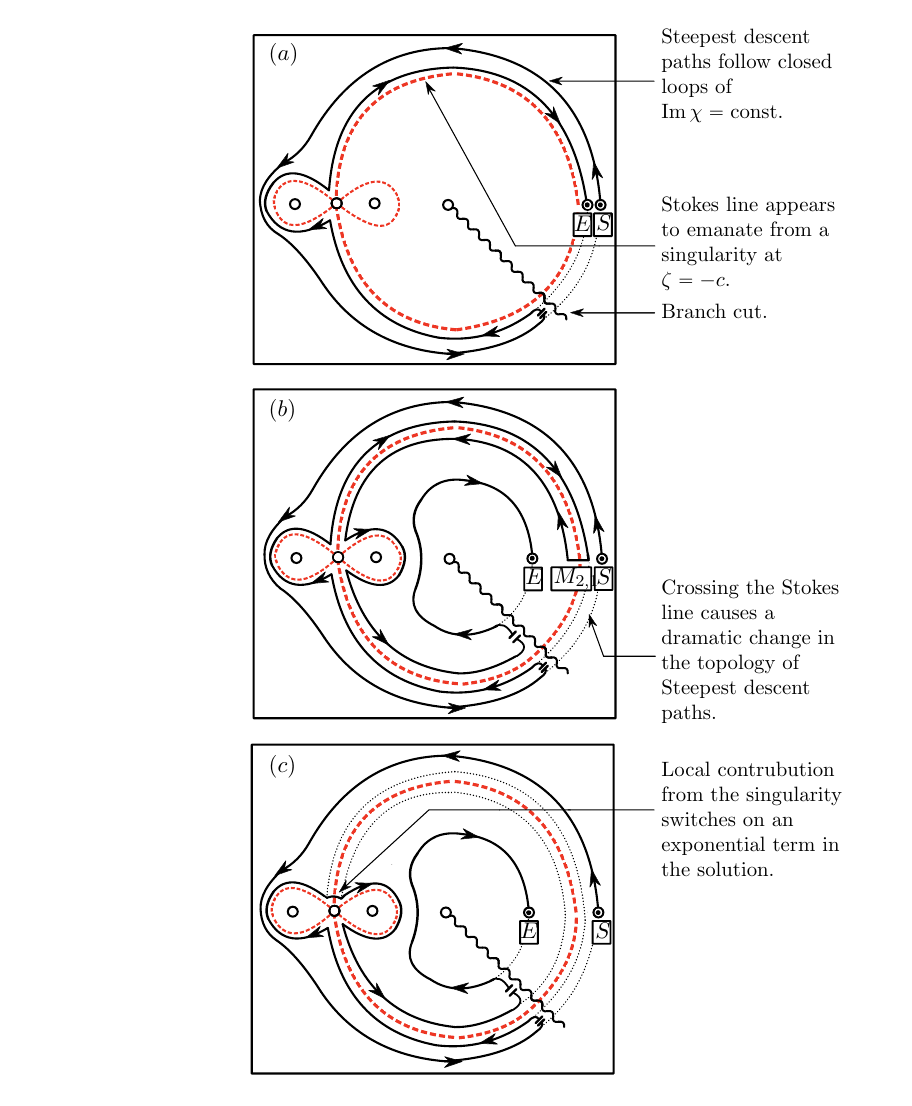}
	
	
	
	
\caption{Illustration of steepest descent paths of \eqref{eqn:mainint} the case of the sharp bump. The integral is initially defined from S to E, and then is deformed along paths of steepest descent (solid black curves) in the direction indicated by arrows. Open circles denote critical points, while wavy lines indicate branch cuts. The integration path continues to deform onto further Riemann sheets upon encircling critical points---this is indicated with a double bar. The Stokes line (red dash) emanates the critical point $\zeta = -c$ and connects with the free surface. When the endpoint $E$ is upstream of the Stokes line (a) the process results in only contributions from the integration endpoints. In contrast, when the endpoint $E$ is upstream of the Stokes line (b,c) the steepest descent paths loop around the critical point C, causing an oscillatory contribution to the asymptotic approximation of the integral solution. \label{fig:sharp-crossing}}
\end{figure}
\clearpage
}

\subsection*{(Case I) Only endpoint contributions}

We first consider the case where the start and end points, $\tilde\zeta = S$ and $\tilde\zeta = \zeta = E$, are chosen sufficiently far upstream so as to lie outside the homoclinics of $C$. This is illustrated in \cref{fig:sharp-crossing}(a). The contour begins from $S$, encircling all four branch points in closed loops, with each full rotation accessing a new Riemann sheet with ever decreasing values of $\Re\chi$. Once in the valley where $\Re\chi\to-\infty$, the contour re-ascends in a clockwise manner following the constant-phase contour of $E$. By the theory of steepest descents, the contributions from points $S$ and $E$ exponentially dominate those from elsewhere on the path. We note also that since the connection between the two paths is made in a valley where $\Re\chi\to-\infty$, any error introduced from the jump between constant-phase contours is exponentially subdominant to \eqref{eqn:sharpcase1}. Thus to leading order, we may loosely write 
\begin{equation}\label{eqn:sharpcase1}
I(\phi) \sim I_\text{endpt.},
\end{equation}
where the right hand side may be evaluated using integration by parts.
\begin{equation}
  \int_{S}^\zeta q^\prime_0(\tilde\zeta)e^{\chi(\tilde\zeta)/\ep}\de \tilde{\zeta} =\ep\frac{q_0(\tilde\zeta)}{\chi^\prime(\tilde\zeta)}\e^{\chi(\tilde\zeta)/\ep}\Big|_S^\zeta-\int_{S}^\zeta \dd{}{\tilde\zeta}\left(\frac{q_0(\tilde\zeta)}{\chi^\prime(\tilde\zeta)}\right)e^{\chi(\tilde\zeta)/\ep}\de \tilde\zeta,
  \end{equation}
where the dashes indicate differentiation with respect to $\zeta$. Care should be taken not to confuse this with the earlier notation in which it denotes differentiation with respect to $w$.
It is now straightforward to take the start point of integration to be upstream infinity, $S\to\infty$. We impose the radiation condition of uniform uniform upstream flow and consequently the contribution from $S$ vanishes since $q^\prime(S) \to0$. Thus we may determine that the dominant contribution to the integral solution \eqref{eqn:mainint} is from an expansion about the point $E$:
  \begin{equation} \label{eq:sharp_basicasym}
  \overline{q}(\zeta)\sim-\im\ep\zeta q_0^3(\zeta)\dd{q_0}{\zeta} +\Oh(\ep^2)
  \end{equation}
where we have used \eqref{eq:q0_sharp_steep} to eliminate the $\chi^\prime$ term. Thus we conclude that for points sufficiently far upstream of the step, there are no free-surface waves present. The above is observed as the $\Oh(\ep)$ term of a typical asymptotic expansion applied to the differential equation \eqref{eqn:N1}.

\subsection{(Case II) Endpoint and exponentially-small contributions}
We now demonstrate that the homoclinic from $C$ which encloses the origin is a Stokes line. Let us choose start and end points, $\tilde\zeta = S$ and $\tilde\zeta = \zeta = E$, which are upstream and downstream of this line respectively. Moreover, we introduce two intermediate points $M_1$ and $M_2$ which are infinitesimally upstream and downstream of the homoclinic respectively. The four points may be seen in \cref{fig:sharp-crossing}(b). Following \cite{trinhTopologicalStudyGravity2016}, we split the integration using the intermediate points, and write the procedure in shorthand as follows: 
\begin{equation}
I(E) = \int_{S}^{E}  = \int_{S}^{M_1}  +\int_{M_2}^{E} 
\end{equation}
For brevity and clarity, we will continue to use the above shorthand, often omitting the integrand quantities.

For the first integration above, both endpoints $S$ and $M_1$ are upstream of the homoclinic and therefore the integral is evaluated as in \emph{Case I} above. For the second integral, the steepest descent contour from $M_2$ repeatedly encircles the origin in an anticlockwise manner before making a jump on sheet $\chi_{(0,0,0,\infty)}$ and re-ascending in a clockwise manner along the constant-phase contour of $E$. Finally, the sections of the adjacent constant-phase paths of $M_1$ and $M_2$ which lie on the primary Riemann sheet $\chi_{(0,0,0,0)}$ sum to zero, and we take $S\to\infty$ leaving us with
\begin{equation} \label{eq:temp123}
I(E) = I_\text{endpt.} + I_\text{exp} = I_\text{endpt.}+\intleft_C.
\end{equation}
The steepest descent paths are seen in \cref{fig:sharp-crossing}(b) and \cref{fig:sharp-crossing}(c). We note that the second term, $I_\text{exp}$, which is symbolically referenced by $\intleft_C$ represents integration in the $\zeta$-plane around point C, and between the local angles of emergence of the left and right homoclinics, $\nu = 7\pi/13$ and $\nu = 15\pi/13$ respectively. It is this contribution which will result in exponentially small waves in the solution \eqref{eqn:mainint}. Specifically, the integral is given by 
\begin{equation} \label{eq:Iexp_temp}
  I_{\text{exp}} = \intleft_C \dd{q_0}{\tilde\zeta} \e^{\chi(\tilde\zeta)/\ep}\de \tilde\zeta. 
\end{equation}
We note that indeed $\zeta = -c$ corresponds to a saddle point of the integral, with $\chi^\prime(-c)  = 0$. Using \eqref{eq:q0_sharp_steep} we have that as $\zeta \to -c$, $q_0$ and $\chi$ behave like
\begin{align}\label{eqn:sharp-q0scaling}
q_0 \sim \Lambda_1 (\zeta+c)^{-2\sigma}\qquad
\text{and}\qquad \chi \sim\Lambda_2(\zeta+c)^{6\sigma+1},
\end{align}
where $\Lambda_1 = (a-c)(b-c)$ and $\Lambda_2 = -\im c^{-1}(6\sigma+1)^{-1}\Lambda_1^{-3}$.
To simplify the exponential, we make the transformation
   \begin{equation}
  u \equiv \frac{\chi(\xi)}{\ep} \sim \frac{\Lambda_2}{\ep}(\zeta+c)^{6\sigma+1}.
  \end{equation}
  Recall that in the $\zeta$ plane, the integral $\intleft_C$, corresponds to local integration around the point $\zeta = -c$ in the shape of an arc. Under the transformation $u = \chi/\ep$, the steepest-descent integration in the $u$-plane then corresponds to a local Hankel contour around $u = 0$, denoted $H_0$ \citep{trinhReducedModelsGravity2017}. This contour originates from $-\infty-0\im$, encircles the origin, and tends to $u = -\infty+0\im$. The integral \eqref{eq:Iexp_temp} therefore becomes
\begin{equation}\label{eqn:sharp-case2form}
  I_{\text{exp}} \sim 2\sigma c\Lambda_1^2\Lambda_2^{\frac{8\sigma+1}{6\sigma+1}}\ep^{1-\frac{8\sigma+1}{6\sigma+1}}
  \int_{H_0} u^{-\frac{8\sigma+1}{6\sigma+1}} \, \e^u \, \de u.
  \end{equation}
To evaluate the integral term, we use the definition of the Gamma function \citep{abram}. Combining the final result with the base asymptotic expansion from the endpoint contributions, we thus have 
\begin{equation}
  \ql \sim \left[-\im\ep\zeta q_0^3(\zeta)\dd{q_0}{\zeta}+\Oh(\ep^2)\right] +\Lambda\ep^{1-\frac{8\sigma+1}{6\sigma+1}}\e^{-\chi(\zeta)/\ep},\label{eqn:SD-form-sharp}
  \end{equation}
where the constant $\Lambda$ is given by
  \begin{equation}
  \Lambda = \frac{4\pi\im\sigma c\Lambda_1^2\Lambda_2^\frac{8\sigma+1}{6\sigma+1}}{\Gamma\left(\frac{8\sigma+1}{6\sigma+1}\right)}.
  \end{equation}
The specific value of the constants that appear in the above value of $\Lambda$ are not important for our purposes.

The important takeaway is that the endpoint $E$ crossing the dashed homoclinic results in a switching-on of an exponentially small oscillatory term. Therefore we may conclude that for points sufficiently far downstream from the sharp bump, the free-surface waves arise from a switching-on of the dominant contribution about $\zeta = -c$ corresponding to the sharp corner at the apex of the bump. Indeed the problem exhibits the Stokes phenomenon across the central homoclinic in \cref{fig:sharp-conts}; this is the Stokes line (or a projection thereof). 

\subsection{Comparisons with numerical simulations} \label{sec:num_sharp}

In this section, we verify the asymptotic predictions above using numerical results of the full boundary-integral governing equations \eqref{eq:govset}. The boundary-integral equations are computed using the numerical scheme described in Chap.~6 of \citet{vanden2010gravity}. More details of the numerical algorithm are found in Chap.~3 of \citet{yyanis_thesis}.

The asymptotic approximation \eqref{eqn:SD-form-sharp} implies that the surface waves have a far-field amplitude of
\begin{equation}\label{eqn:exp-mag-sharp}
\sim \left|\Lambda\ep^{1-\frac{8\sigma+1}{6\sigma+1}}\e^{-\chi(\zeta)/\ep}\right| = |\Lambda|\ep^{1-\frac{8\sigma+1}{6\sigma+1}}\e^{-\Re\chi(\zeta)/\ep},
\end{equation}
Since the $N=1$ truncation model is not expected to accurately predict the prefactor, we are not concerned with verifying the prefactor value. The singulant, $\chi$, is calculated from \eqref{eqn:N1chi} with $\zeta^* = -c$. Since the leading-order velocity $q_0$ is purely real on the free surface, then $\Re\chi$ is constant on the free surface and it is sufficient to calculate the integral to any convenient point on the free surface. For example, for $\sigma = 3/8, a = 1$, and $b = 2$, a standard numerical quadrature using \emph{e.g.} the software Mathematica gives $\Re\chi\approx 2.7250$ on the free surface. 

\begin{figure}[htb]
\centering 
\includegraphics[scale=0.9]{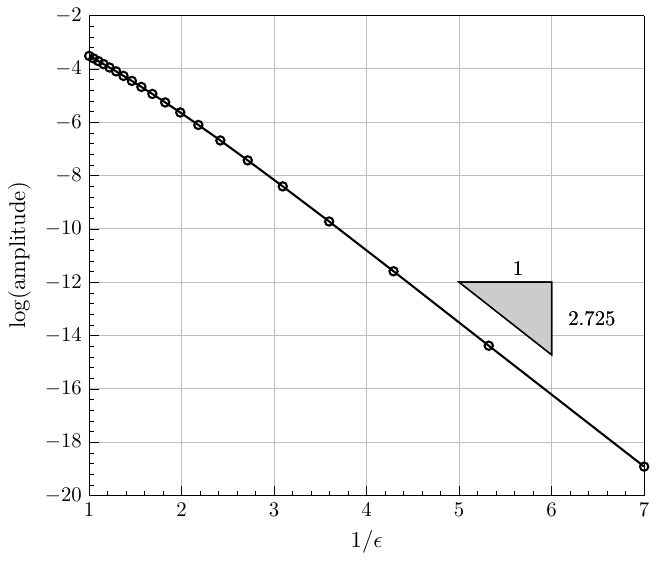}
\caption{Comparison between numerical wave amplitudes (circles) calculated for the full water-wave equations and the analytical gradient predictions from \eqref{eqn:exp-mag-sharp} for the sharp bump. The sharp bump geometry is prescribed by \eqref{eq:q0_sharp_steep} with $a = 1$, $b = 2$, and $\sigma = 3/8$. Numerical wave amplitudes are calculated by a boundary-integral finite difference scheme with $N = 1000$ mesh points distributed over $-30 \leq \phi \leq 30$. The gradient value of $-\Re\chi \approx -2.725$ is computed via integration of \eqref{eqn:N1chi}. \label{fig:compare-sharp}}
\end{figure}

In \cref{fig:compare-sharp}, we compare the magnitude of the waves predicted by the theory of exponential asymptotics, \eqref{eqn:exp-mag-sharp}, with the amplitude of the numerical solution to the full boundary-integral problem \eqref{eq:govset}. The agreement of the exponential trend (gradient) on the semilog plot is excellent, and improves as $\ep\to 0$. 

%


\section{Steepest descents for the smooth bump}\label{sect:smooth-sd}

We now study the integral \eqref{eqn:mainint} for the case of the smooth bump \eqref{eqn:smoothed-qs}, repeated here,
\begin{equation}
q_0(\zeta) = A\left(\frac{\zeta+1}{\zeta+b}\right)^{B(\zeta+c)},
\end{equation}
where $A = e^{2\sigma}$ and $B = \frac{2\sigma}{(b-1)}$.
 We will focus on the case in which $b = 2$ and $\sigma = 3/8$. We note that $\sigma > 1/3$ is such that the bump endpoints do not generate and Stokes lines into the fluid and therefore any present waves are associated solely with the smooth bump.

\begin{figure}[htb]
\centering
\includegraphics{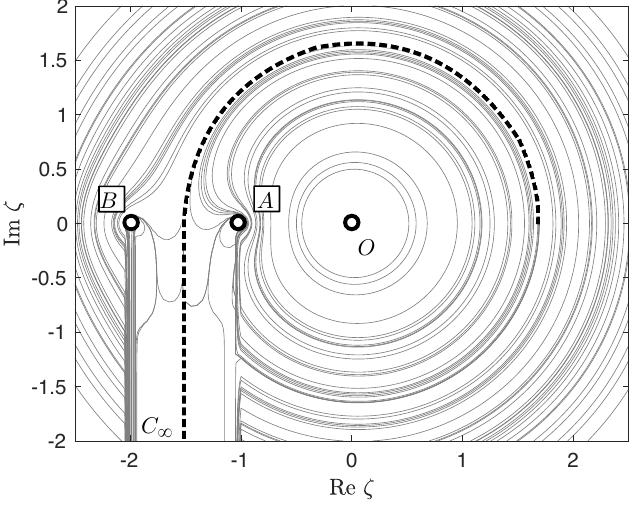}
\caption{A contour plot of $\Im\chi$ in the case of the smooth bump problem. The bump has corners located at $A = 1, B = 2$ and emerges at an angle $\vartheta = 3\pi/8$ to the horizontal (see \cref{fig:geoms}). Here $\chi$ and is obtained according to \eqref{eqn:N1chi-smooth}. The free surface and fluid bottom are represented by the positive- and negative-real axes respectively, while upper-half-plane is associated with the inner-fluid region. In contrast to the three homoclinics in Fig.\ref{fig:sharp-conts}, there is a curve (dashed) which appears to emanate from $\zeta = -b -\im\infty$, travel through the fluid region, and intersect the free surface. As we demonstrate later, this is the Stokes line associated the switching-on of downstream ripples. A selection of constant-phase paths, $\Im\chi = \text{const.}$ is indicated by grey homoclinics. As we demonstrate later, steepest descent paths lie along these curves.\label{fig:smooth-conts}}
\end{figure}

A selection of level sets of $\Im\chi$ is shown in \cref{fig:smooth-conts}. These contours form closed loops, according to a Riemann structure topography which is induced by the three critical points at $B, A,$ and the origin. Points $A$ and $B$ correspond to the bottom corners of the smooth bump, while the logarithmic singularity at the origin is an artefact of the mapping $\zeta = \e^{-w}$ from the potential plane to the $\zeta$-plane. The logarithmic singularity is of little consequence in the following analysis. In contrast to the sharp bump structure seen in \cref{fig:sharp-conts}, there is no critical point between $B$ and $A$. However, there appears to be singular behaviour as $\Im\zeta \to -\infty$ with $-b<\Im\zeta<-1$. 


We follow the approach of the previous section and consider two distinct cases in which the endpoint of the integral \eqref{eqn:mainint} lies upstream (\emph{Case I}) and downstream (\emph{Case II}) of the red dash line in \cref{fig:smooth-conts}. We will demonstrate that in an analogous manner to the problem of the sharp bump, a transition from \emph{Case I} to \emph{Case II} causes a switching on of an exponentially-small oscillatory term in the asymptotic approximation to the integral. Curiously, in contrast to the sharp bump, the critical line (the projected Stokes line) does not seem to emanate from any critical point in the physical geometry; rather it is generated by an essential singularity lying at infinity.

%

\emph{On the singularities.} First, we take note of the critical points of the singulant, $\chi$. As with the sharp bump, there are branch points at $A$ and $B$ corresponding to the bottom corners of the bump. We choose branch cuts vertically downwards from these points, and choose the logarithmic branch cut from the origin to extend diagonally into the fourth quadrant. In contrast to the sharp bump there is no singular point associated with a bump apex. We reuse the index notation of the previous section to denote the individual Riemann sheets, $\chi_{(k_1 ,k_2 ,k_3)}$, with the primary sheet $\chi_{(0,0,0)}$ taken as that which contains the physical fluid, and where the indexing numbers $k_i\in\mathbb{Z}$ indicate the number of anticlockwise rotations around the branch points $\zeta= \{-b, -a, 0\}$ needed to access the sheet from $\chi_{(0,0,0)}$. The primary Riemann sheet is shown in \cref{fig:sharp-conts}.

There is unbounded behaviour exhibited at infinity which is seen if we take $\Im\zeta\to-\infty$ in a certain manner. To see this, set $\zeta = \zeta_r + i\zeta_c$, where $-b<\zeta_r <-1$, and specify the choice of branch by setting 
$\Arg(\zeta+1)\sim3\pi/2$ and $\Arg(\zeta+b)\sim-\pi/2$, valid in this limit. 
This is equivalent to starting in the upper-half-plane of the primary sheet between $A$ and $B$ and travelling vertically downwards to $\zeta = \zeta_r-\im\infty$. The behaviour of $q_s = q_0$ is then given by
\begin{equation}
	q_0 \sim A e^{-2\pi B\zeta_c}e^{-2\pi\im B(\zeta_r+c)}\exp\left[B(\zeta+c)\log\left|\frac{\zeta+1}{\zeta+b}\right|\right]. 
\end{equation}
Consequently we see that $|q_0| \to \infty$ as $\zeta_c \to -\infty$. Moreover, there is a series of peaks and troughs depending on the oscillatory factor, \emph{i.e.} the value of $\zeta_r$. If we let $\tilde{k} = 2\pi B$ then we may write $q_0 \sim A \e^{\i\tilde{k}\zeta}$ as $\zeta \to -c-\im\infty$ confirming the oscillatory behaviour.

 \Cref{fig:smooth-conts} illustrates that in this far-field limit, the constant-phase paths are roughly vertical. We can infer the presence of a partitioning line which joins the fluid bottom with the free surface, indicated by a red dashed line in \cref{fig:smooth-conts}. This is reminiscent of the Stokes line in the previous problem (\emph{cf.} the central homoclinic in \cref{fig:sharp-conts}). However this curve is not a homoclinic, and appears to extend vertically downwards from $\zeta = -c$, presumably to the singularity at $\zeta = -c-\im\infty$, which we denote as $C_\infty$. Thus in this case, we should define the point of origin, $\zeta^*$ in \eqref{eqn:N1chi} as $C_\infty \approx -c - \im\infty$, i.e. 
 \begin{equation}\label{eqn:N1chi-smooth}
\chi(\zeta) = \int_{C_\infty}^{\zeta}\frac{\im}{\tilde\zeta q_0^3(\tilde\zeta )}\de \tilde{\zeta}.
 \end{equation}
Numerical quadrature of the above integral, for various specific cases of $q_0$, will be used in \cref{sec:num_bump} for the comparisons to the full water-wave problem.

In the same manner of the analysis in \cref{sect:sharp-sd}, we now examine the two distinct cases \emph{Case I} and \emph{Case II}.

\subsection*{(Case I) Only endpoint contributions}

When the endpoints, $\tilde\zeta = S$ and $\tilde\zeta = E$, both lie upstream of the dashed line, the asymptotic evaluation of \eqref{eqn:mainint} is effectively the same as for Case I for the sharp bump (\emph{cf.} \cref{sect:sharp-sd}). The starting point of integration is taken to be upstream infinity, $S\to\infty$ whereby the contribution from $S$ vanishes due to required uniform flow upstream. Then, the asymptotic expansion about the $E$ produces \eqref{eq:sharp_basicasym} and this re-derives the standard asymptotic expansion of \eqref{eqn:N1}. The free-surface is effectively wave-free on this portion of the free surface.

\afterpage{
\begin{figure}[htb]
\includegraphics[scale=1]{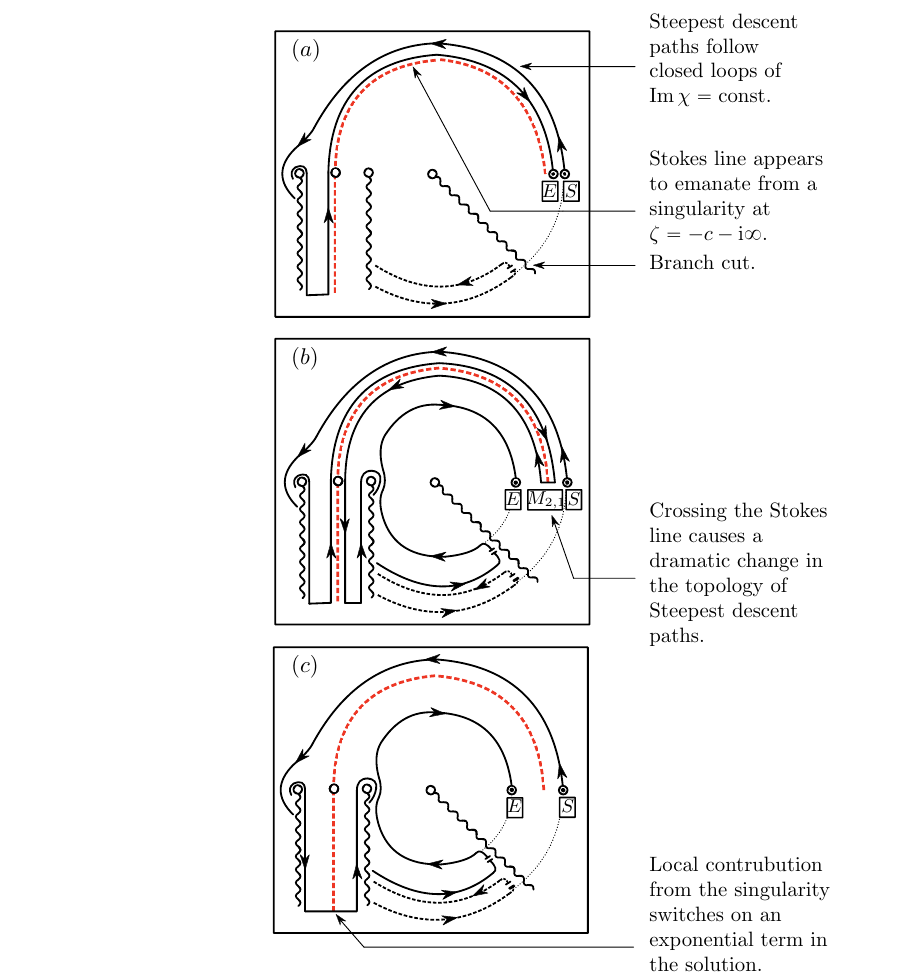}
	
	
\caption{Illustration of the steepest descent deformation for the integral \eqref{eqn:mainint} for flow over a smooth bump, as shown in the $(\Re\zeta, \Im\zeta)$-plane. The initial path of integration runs from S to E (black circles); the deformed integrals along steepest descent/ascent paths are shown with the black arrows. Open circles denote critical points, while wavy lines indicate branch cuts. The Stokes line (red dash) appears to emanate from a point $\zeta = -c-\im\infty$. When the endpoint $E$ is upstream of the Stokes line (a) the process results in only contributions from the integration endpoints. However, when the endpoint $E$ lies upstream of the Stokes line (b,c) the steepest descent paths interact with the essential singularity at infinity, causing an oscillatory contribution appear. \label{fig:smooth-crossing}}
\end{figure}
\clearpage
}

\subsection*{(Case II) Endpoint and exponentially-small contributions}

We now demonstrate that the red dash line in \cref{fig:smooth-conts} connecting $\zeta = C_\infty$ to the free surface is a Stokes line for this flow. We choose the endpoints of integration, $\tilde\zeta = S$ and $\tilde\zeta = \zeta = E$, on the free surface, upstream and downstream respectively of this line. We split the integration using two intermediate points, $M_1$ and $M_2$, which are on the free surface infinitesimally upstream and downstream of the dashed line respectively. Then schematically:
\begin{equation}
I(E) = \int_{S}^{E} = \int_{S}^{M_1}+\int_{M_2}^{E}.
\end{equation}
The first integral from $S$ to $M_1$ may be evaluated as in \emph{Case I} since both endpoints are upstream of the red dash line. The path of steepest descent for the second integral, from $M_2$ to $E$, is illustrated in \cref{fig:smooth-crossing}(b) and behaves as follows.  We start from $M_2$ and follow the steepest descent contour, \emph{i.e.} the level set $\Im\chi(\zeta) = \Im\chi(M_2)$. This runs alongside the red dashed curve (Stokes line), descending to $C_\infty$ before re-ascending and winding clockwise around $\zeta = -a$. From here, it repeatedly encircles the origin in an anticlockwise manner, descending into a valley of $\Re\chi$ before making the jump to the level set 
 $\Im\chi(\zeta) = \Im\chi(E)$. This connection is made on sheet $\chi_{(0,0,\infty)}$. Finally, the path unwinds around the origin in a clockwise direction and joins with the endpoint $E$. 
 
  Analogously to the technique used in the sharp bump analysis, the section of the paths from $M_1$ and $M_2$ which lie on the primary sheet sum to zero. This produces the final contour sketched in \cref{fig:smooth-crossing}(c), and admits contributions written in the schematic form
\begin{equation}
I(E) \sim I_\text{endpt.}+I_\text{exp} = \int_E +\int_{C_\infty},
\end{equation}
where the final term denotes a contribution associated the singularity at infinity, $C_{\infty}$, and again we have taken $S \to \infty$ to apply the upstream radiation condition. We briefly outline the difficulties in explicitly evaluating the second integral contribution near $C_\infty$. 

Firstly, the integral may be written 
\begin{equation}\label{eqn:smooth-osc-int}
I_{\text{exp}} = \int_{C_\infty} \dd{q_0}{\tilde\zeta} \, \e^{\chi(\tilde\zeta)/\ep}\de \tilde\zeta, 
\end{equation}
and consists of consideration of a steepest descent path from $C_\infty$. In the typical flow configurations (such as in the sharp bump) the leading-order speed, $q_0$, observes a power-law scaling near the critical point [\emph{cf.} equation \eqref{eqn:sharp-q0scaling}]. However, in this problem we are dealing with an essential singularity with exponential blow-up, $q_0 \sim A e^{i\tilde{k}\zeta}$ as $\zeta \to -c+\im\infty$, where $\tilde{k} = 2\pi\sigma/(b-a)$ is a positive real number. This implies that the singulant, $\chi$, scales as
\begin{equation}
\chi\sim\Ei(-3\im\tilde k \zeta)\sim\frac{\im\e^{-3\im\tilde k \zeta}}{\tilde k \zeta} \quad \text{as} \quad \zeta \to -c+\im\infty,
\end{equation}
where $\Ei$ denotes the exponential integral.

In a more regular problem such as the sharp bump, the integration path in \eqref{eqn:smooth-osc-int} is typically re-posed, via a re-scaling, to a Hankel contour by means of a transformation such as $u \sim \chi(\tilde\zeta)/\ep$ near the critical point. In this problem we may therefore expect to be able to use the transformation
$u = \im\e^{-3\im\tilde k \zeta}/(\tilde k \zeta)$. However, the inability to analytically invert the relationship causes difficulty with performing the necessary local analysis of the integral. The authors were unable to develop a simpler expression for the local steepest descent contribution of \eqref{eqn:smooth-osc-int}. Notwithstanding the inability to approximate the integral contribution explicitly, there is strong numerical evidence that the contribution does produce the required exponentially-small oscillations as $\ep \to 0$; this is presented in the next section.

\subsection{Comparisons with numerical simulations} \label{sec:num_bump}

We conjecture that as $\ep\to0$, downstream waves produced by flow over the smooth bump have magnitudes that scale with $\mathcal{A}e^{-\Re\chi/\ep}$. This is then predicted by the scaling on the local integral contribution of \eqref{eqn:smooth-osc-int} due to the essential singularity, $C_\infty$. 

\begin{figure}[htb]
\centering 
  \includegraphics[scale=0.9]{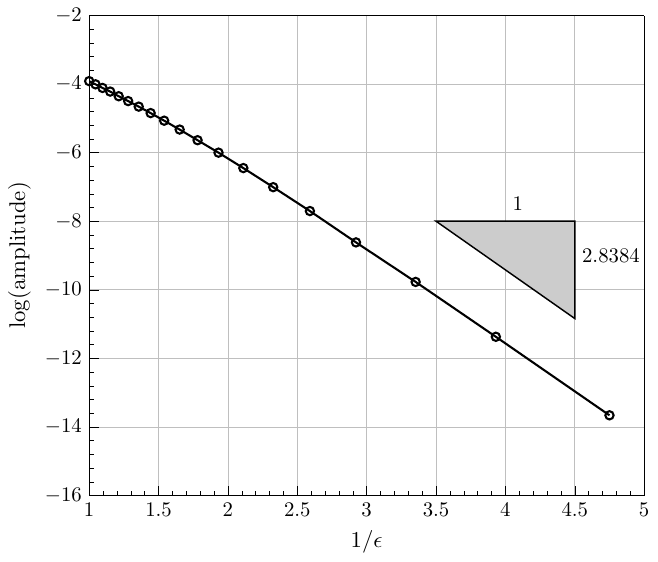}
\caption{Comparison between numerical wave amplitudes and analytical gradient predictions for the smooth bump with $a = 1$, $b = 2$, and $\sigma = 3/8$. The numerical solutions correspond to the full boundary-integral equations, computed using a finite-difference mesh of $N = 700$ points distributed over $-25 \leq \phi \leq 30$. The gradient value of $-\Re\chi \approx -2.8384$ is calculated from numerical quadrature of \eqref{eqn:N1chi} with initial point of integration $C_\infty$. \label{fig:compare-smooth}}
\end{figure}

Like in the previous \cref{sec:num_sharp}, we calculate the solution of the full boundary integral equations \eqref{eq:govset} using a standard finite-difference scheme [cf. additional details in \cite{yyanis_thesis}], now with the shape function specified by \eqref{eqn:smoothed-qs}. In \cref{fig:compare-smooth}, we present numerical amplitude values for those downstream waves corresponding to geometrical parameters set to values of $a = 1$, $b = 2$, $\sigma = 3/8$. Based on the image and fit to the asymptotically predicted gradient of $\e^{-\Re(\chi)/\ep}$, the evidence seems to confirm the exponential smallness of the free-surface waves for the smooth bump problem.

\begin{figure}[htb]
\centering 
  \includegraphics[scale=0.9]{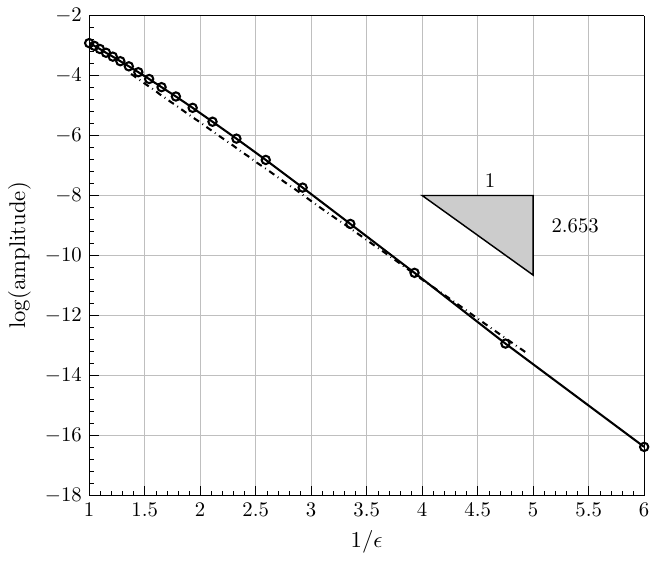}
\caption{Comparison between numerical wave amplitudes and analytical gradient predictions for the smooth bump with $a = 1$, $b = 3$, and $\sigma = 1/2$. The circles denote numerical solutions to the full boundary-integral equations using a mesh of $N = 700$ points distributed over $-25 \leq \phi \leq 30$. The gradient appproximation of $-\Re\chi \approx -2.653$ is calculated from numerical quadrature of \eqref{eqn:N1chi} with initial point of integration $C_\infty$. These results compare favourably with the corresponding amplitude curve (shown dashed) extracted from fig.~10 of \cite{pethiyagodaEfficientComputationTwodimensional2018}.\label{fig:compare-smooth2}}
\end{figure}

A similar plot is given in \cref{fig:compare-smooth2} for the case of $a = 1$, $b = 3$, and $\sigma = 1/2$, which was studied by the numerical work of \cite{pethiyagodaEfficientComputationTwodimensional2018}. In this case, the computed value of $\Re\chi \approx 2.65$, calculated by integrating \eqref{eqn:N1chi-smooth} matches well with the fitted value of $\approx 2.63$ seen in figure~11 of that work. We can verify similar agreement between their fitted values and our value calculated via \eqref{eqn:N1chi-smooth}; for instance changing to $b = 10.5$, their fitted value is $\approx 2.46$ as compared to the numerical quadrature yielding $\Re\chi \approx 2.48$.


%
 

\section{Discussion}

Our original motivation was to examine the apparently simple scenario of flow over a smooth bump, as visualised in \cref{fig:geoms}(b). The criterion by \cite{chapmanExponentialAsymptoticsGravity2006} indicates that when a corner point within a potential flow presents an in-fluid angle greater than $2\pi/3$, a Stokes line emerges. Across such a Stokes lines, an exponentially-small wave contribution is switched on. The case of the smooth bump does not invalidate their original statement [as quoted below \eqref{eq:earlysing}]---however, it should be emphasised that there exists other classes of relevant flows where waves are not attributed to the standard singularities of potential flow (algebraic branch points or logarithms). Moreover, such geometries may even be considered to be well-behaved: the smooth bump presented in this work is one such example.

By using a model that reduces the full boundary-integral framework into an integral form \citep{trinhReducedModelsGravity2017}, we have been able to demonstrate that the flow past the particular smooth bump does indeed have a Stokes line that intersects the free surface. However, unlike previously studied cases of exponential asymptotics in free-surface flows [\emph{e.g.} \citep{chapmanExponentialAsymptoticsCapillary2002,chapmanExponentialAsymptoticsGravity2006,trinhWavelessShipsExist2011,trinhWakeTwodimensionalShip2014,lustriFreeSurfaceFlow2012}] the Stokes line is unusually attributed to the presence of an essential singularity in the analytic continuation of the free surface (rather than an obvious geometrical feature of the bottom topography). 

There are a number of open challenges this study motivates. First, despite the authors' best attempts, we were unable to develop the exponential asymptotics of the smooth bump problem to greater detail: for instance, we were unable to derive the full analytical form of exponential switching, including the details of the prefactor [as in \eqref{eqn:bump-exp-terms}]. Numerical evidence suggests that the wave-amplitude is exponentially small (to leading-order); however, it is possible that {\itshape e.g.} logarithmic factors of $\log\ep$, render comparisons to numerical results at standard levels of precision challenging and misleading [cf. the example in sec.~{6.3} of \citealt{crewResurgentAspectsApplied2023}].

The analysis near the essential singularity seems highly challenging, even for the case of the simplified truncation model presented in \cref{sect:exp-asy-review}. It might be the case that other water-wave formulations, such as those that work directly with physical coordinates rather than potential-plane variables \citep{ablowitz2006new} present alternative or advantageous formulations for studying flows past smooth geometries. However, this does not alleviate all of the issues highlighted in this work, which is that intrinsically, exponential asymptotics relies upon the study of singularities in the analytic continuation of the solutions---however, analytic continuation is ill-posed and it is possible to achieve quite `wild' singularity behaviours with minor modifications of the associated flow geometry.

A natural question is whether the ``unusual" Stokes line we have discovered in our bump problem, and the pathological behaviour in the far field, is typical of smooth geometries or simply a result of our particular geometry specification. In a more general version of the problem, the specified $\theta$ for the channel bottom might be such that the corresponding shape function from \eqref{govset_int} is not known analytically, and thus requires numerical calculation. It is interesting to consider whether an exponential asymptotics framework can be designed for such generality.

There seems to be a deep connection between this problem, and the exponential asymptotics problem studied by \cite{akylas1995short}. In their work, the authors study the steady forced Korteweg-de Vries equation, 
\begin{equation} \label{eq:fkdv}
    \mu^2 \dd{^2 u}{x^2} + u - \ep u^2 = f(x),
\end{equation}
on an infinite domain with $u \to 0$ as $x \to \infty$ and $f \to 0$ as $x \to \pm \infty$. In the limit $\mu, \ep \to 0$, the solutions exhibit exponentially-small oscillations. The authors present the cases of $f(x) = \sech(x)$, $\sech^2(x)$, and $\e^{-x^2}$. The first two cases are associated with typical singularities (poles) in the analytic continuation of the $\sech(x)$ and $\sech^2(x)$ functions. However, they note that the case of the entire Gaussian function $\e^{-x^2}$ is extremely challenging. Like our problem, there is an essential singularity as $|x|\to\infty$. Their approach to the exponential asymptotics leverages a spectral scheme applied to the Fourier transform of \eqref{eq:fkdv}; with great difficulty, the authors were able to derive the exponentially-small contributions to $u$ in this special case. Our problem, however, either for the full boundary-integral equation \eqref{eq:govset} or the simplified formulation \eqref{eqn:trinh2017-reduced}, is not so easily studied via Fourier methods.

This question of the development of beyond-all-orders techniques for problems with more general singularity structures remains important. We are encouraged, in particular, by new results that have applied the above spectral approaches to more challenging settings \citep{kataoka2023nonlinear}, including the companion paper published in this same Special Issue \citep{kataoka2023WAWA}. The inclusion of additional physical effects such as capillarity \citep{shelton2022exponential,lustriThreedimensionalCapillaryWaves2019} and vorticity \citep{shelton2023exponential} also presents many intriguing cases for study.

\mbox{}\par
\noindent \textbf{Acknowledgements.} The authors thank Jon Chapman (Oxford), John King (Nottingham), and Samuel Crew (Bath) for many interesting and useful discussions surrounding the current work. The authors would like to thank the Isaac Newton Institute for Mathematical Sciences for support and hospitality during the programme Applicable Resurgent Asymptotics when work on this paper was undertaken. This work was supported by EPSRC Grant Number EP/R014604/1. PHT gratefully acknowledges support from EPSRC Grant Number EP/V012479/1. YJ-M was supported by a scholarship from the EPSRC Centre for Doctoral Training in Statistical Applied Mathematics at Bath (SAMBa), under the project EP/S022945/1.

\mbox{}\par
\noindent \textbf{Declarations.} On behalf of all authors, the corresponding author states that there is no conflict of interest. The results of the manuscript are generated using standard routines and the manuscript has no associated data.


\end{document}